\documentclass[journal,draftcls,onecolumn,letterpaper,twoside,11pt]{IEEEtran}

\usepackage[cmex10]{amsmath}
\usepackage{latexsym,amssymb,cite,bbm,epsf,amsthm,subfigure,algorithm,algorithmic}
\usepackage{graphicx}
\usepackage{enumerate}
\usepackage{mathrsfs}
\usepackage{epstopdf}
\usepackage{bbm}
\usepackage{color}

\newcommand{\bc}{\begin{center}}
\newcommand{\ec}{\end{center}}
\newcommand{\be}{\begin{equation}}
\newcommand{\ee}{\end{equation}}
\newcommand{\ba}{\begin{array}}
\newcommand{\ea}{\end{array}}
\newcommand{\bea}{\begin{eqnarray}}
\newcommand{\eea}{\end{eqnarray}}
\newcommand{\bal}{\begin{align}}
\newcommand{\eal}{\end{align}}
\newcommand{\ei}{\end{itemize}}
\newcommand{\bi}{\begin{itemize}}
\newcommand{\bfi}{\begin{figure}}
\newcommand{\efi}{\end{figure}}
\newcommand{\MB}{\left[\begin{array}}
\newcommand{\ME}{\end{array}\right]}
\newcommand{\nn}{\nonumber}

\newtheorem{thm}{Theorem}

\newtheorem{cor}{Corollary}
\newtheorem{lem}{Lemma}

\newcommand{\Exp}{\mathsf{E}}
\newcommand{\bExp}{\bar{\mathsf{E}}}
\newcommand{\Pro}{\mathsf{P}}

\newcommand{\cT}{\mathcal{T}}
\newcommand{\cH}{\mathcal{H}}

\newcommand{\cN}{\mathcal{N}}
\newcommand{\cI}{\mathcal{I}}
\newcommand{\cX}{\mathcal{X}}

\newcommand{\bN}{\mathbb{N}}
\newcommand{\bR}{\mathbb{R}}
\newcommand{\bC}{\mathbb{C}}

\newcommand{\ind}[1]{\mathbbm{1}_{\{#1\}}}   
\newcommand{\ignore}[1]{{}}

\usepackage{tikz}

\begin{document}

\title{Sequential Decentralized Parameter Estimation under Randomly Observed Fisher Information}

\author{Yasin~Yilmaz\IEEEauthorrefmark{2}\footnote{\IEEEauthorrefmark{2}Electrical Engineering Department, Columbia University, New York, NY 10027.}\;\;
        and \, Xiaodong Wang\IEEEauthorrefmark{2}}
\ignore{\thanks{Y. Yilmaz and X. Wang are with the Department
of Electrical Engineering, Columbia University, New York,
NY, 10027 USA.}
\thanks{G. Moustakides is with the Dept. of Electrical \& Computer Engineering, University of Patras, 26500 Rion, Greece.}}

\maketitle

\begin{abstract}
We consider the problem of decentralized \textcolor{blue}{scalar parameter estimation} using wireless sensor networks with Gaussian noise. Specifically, we propose a novel framework based on level-triggered sampling, a non-uniform sampling strategy, and sequential estimation. The proposed estimator can be used as an asymptotically optimal fixed-sample-size decentralized estimator when the observed Fisher information, i.e., Fisher information without expectation, is deterministic, as an alternative to the one-shot estimators commonly found in the literature. It can also be used as an asymptotically optimal sequential decentralized estimator when the observed Fisher information is random. We show that the optimal centralized estimator under Gaussian noise, which is the maximum likelihood estimator (MLE), is characterized by two processes, namely the observed Fisher information $U_t$, and the observed correlation $V_t$. It is noted that $V_t$ is always random even when $U_t$ is not. In the proposed scheme, each sensor computes its local random process(es), and sends a single bit to the fusion center (FC) whenever the local random process(es) pass(es) certain predefined levels. The FC, upon receiving a bit from a sensor, updates its approximation to the corresponding global random process, and accordingly its estimate. The sequential estimation process terminates when $U_t$ (or the approximation to it) reaches a target value.
We provide an asymptotic analysis for the proposed estimator and also the one based on conventional uniform-in-time sampling under both deterministic and random $U_t$; and determine the conditions under which they are asymptotically optimal, consistent, and asymptotically unbiased. Analytical results, together with simulation results, demonstrate the superiority of the proposed estimator based on level-triggered sampling over the traditional decentralized estimator based on uniform sampling.
\end{abstract}


\section{Introduction}
\label{sec:intro}

Decentralized parameter estimation is a fundamental signal processing task that can be realized in wireless sensor networks. Due to the stringent bandwidth and energy requirements imposed by sensors it is typically performed under the constraints of low bandwidth usage and low communication rate, unlike centralized estimation. That is to say, sensors need to infrequently communicate to the fusion center (FC) in an FC-based network (to the neighboring sensors in an ad hoc network similarly), consuming low bandwidth, e.g., sending only a few bits each time. \textcolor{blue}{In this paper, we propose a novel sequential framework based on MLE for decentralized signal amplitude estimation, a scalar parameter estimation problem.}

\textcolor{blue}{Signal amplitude estimation for wireless sensor networks was studied in a variety of existing works, e.g., \cite{Ribeiro06,Ribeiro06-2,Xiao05,Li07,Luo05,Aysal08,Chen10,Banavar12,Tepedelen11,Banavar10,Tepedelen10,Luo05-2,Xiao06,Papa01,Abdallah01}, considering only the effect of real additive noise. This traditional line of research, from communications perspective, fails to account for the fading channel effect and the quadrature modulation techniques, such as quadrature amplitude modulation (QAM), phase-shift keying (PSK), and minimum-shift keying (MSK), which are the most commonly used techniques in practice. In a typical communications example, sensors observe complex, i.e., quadratic, signals through fading channels.
This motivates us to study, under a sequential framework, estimating the amplitude of a complex signal under fading channels, i.e., with complex random scaling coefficients and complex noise. From a general perspective, our motivation is to sequentially estimate the amplitude of a complex random signal observed under complex noise. Beside these practical aspects our major motivation in this work is to provide asymptotically optimal sequential decentralized estimators. The proposed estimators under a novel sequential framework and their asymptotic performance analysis constitute the main contribution of this work.}

\textcolor{blue}{The problem of vector parameter estimation was also extensively studied in the literature. For example, \cite{Msechu12,Xiao08,Schizas08,Lopes07,Li10,Chouvardas11,Cattivelli10} considered the linear system identification task, where an unknown vector that characterizes the system to be identified is estimated using a linear model with known regressors, i.e., input to the system. More specifically, \cite{Schizas08,Lopes07,Li10,Chouvardas11,Cattivelli10} proposed adaptive estimators which are sequentially updated as new observations are available. In these adaptive estimators, regressor vectors are random and observed by sensors at each time, resembling the fading channel gains in our setup.
The references \cite{Schizas08,Lopes07,Li10,Chouvardas11,Cattivelli10} assume ad hoc sensor networks, whereas \cite{Ribeiro06,Ribeiro06-2,Xiao05,Li07,Luo05,Aysal08,Chen10,Banavar12,Tepedelen11,Banavar10,Tepedelen10,Luo05-2,Xiao06,Papa01,Abdallah01,Msechu12,Xiao08} assume FC-based networks. In the latter group of papers,
sensors use either digital or analog transmission to send their observations to the FC, whereas in the former group only \cite{Schizas08} considered a practical transmission method using quantization.}

In order to conform to the low bandwidth requirement sensors either quantize their observations with a small number of bits, such as 1 bit, (e.g., \cite{Ribeiro06,Luo05,Dabeer06}) or appropriately pulse-shape their analog transmissions (e.g., \cite{Xiao08,Banavar10,Tepedelen10}). Quantization with a small number of bits causes the observations to be recovered in a coarse resolution at the FC, although it is much easier to implement than analog transmission.
Dithering is used in \cite{Dabeer06} to reduce the bias and improve the consistency of a quantization-based-estimator. In \cite{Papa01}, it is shown that random dithering can significantly reduce the Cramer-Rao lower bound (CRLB) compared to the no dithering case. Moreover, in \cite{Balkan10}, deterministic dithering is shown to be optimal in terms of minimizing the CRLB.

In an FC-based network, various types of reporting channels between sensors and the FC have been analyzed in the literature. For instance, \cite{Ribeiro06,Ribeiro06-2,Luo05,Luo05-2,Xiao06,Papa01,Msechu12,Xiao05,Li07,Balkan10} assume orthogonal (parallel) error-free channels; \cite{Lopes07,Li10,Chouvardas11,Cattivelli10} assume orthogonal error-free infinite-bandwidth channels; \cite{Aysal08} assumes orthogonal non-fading continuous channels; \cite{Chen10,Abdallah01,Dabeer06} assume orthogonal discrete channels (BSC); \cite{Banavar12,Tepedelen11} assume non-fading multiple access channel (MAC); and finally \cite{Xiao08,Banavar10,Tepedelen10,Mergen06} assume fading MAC.
In this paper, we will assume orthogonal error-free channels to focus on the proposed novel sequential framework for decentralized estimation, which is described and analyzed in the following sections.
A common assumption among the existing works is the identically distributed noise or noise with same statistics, e.g., \cite{Ribeiro06,Ribeiro06-2,Msechu12,Xiao05,Li07,Luo05,Aysal08,Chen10,Dabeer06,Banavar12,Tepedelen11,Xiao08,Banavar10,Tepedelen10,Mergen06,Luo05-2} except for \cite{Xiao05}, \cite{Li07} and \cite{Tepedelen10}. In this paper, we avoid making such an assumption.

Some of the decentralized estimators proposed in the previous works are universal in the sense that they do not depend on the probability density function (pdf) of the noise at sensors, e.g., \cite{Xiao05,Luo05,Tepedelen10,Luo05-2}. Here we consider a specific noise pdf, namely the Gaussian noise, and maximum likelihood estimator (MLE), which is the optimum estimator and corresponds to the least squares estimator (LSE) in that case. LSE can be used as a universal estimator under any noise distribution.

\textcolor{blue}{The estimators in \cite{Xiao05} and \cite{Luo05-2} are also independent of the network size and the sensor index}, i.e., robust to changes in the network size (sensor addition/failure), which is a practically desired feature for decentralized estimators \cite{Xiao06}. Similarly, our estimators are robust in that sense (although different thresholds are assumed at each sensor as a general case, the derivations and analysis also cover the specific case of using the same threshold). Most of the estimators in the literature, including the ones in the references above, except for \cite{Xiao05} and \cite{Luo05-2} as already noted, are not robust in that sense.

All of the references above, except for \cite{Papa01,Schizas08,Lopes07,Li10,Chouvardas11,Cattivelli10}, perform fixed-sample-size (one-shot) estimation. However, as stated in \cite{Papa01}, it is not possible in fixed-sample-size estimation to further refine the quality of the estimate before and after the estimation time, unlike the sequential estimation. Moreover, it is natural to expect that sequential estimators require significantly less number of samples than their fixed-sample-size counterparts to achieve the same quality of estimate, as it is known that sequential detection methods, on average, requires approximately four times less samples than their fixed-sample-size counterparts for the same level of confidence \cite[Page 109]{Poor94}. Hence, in this paper we are interested in sequential decentralized estimators rather than fixed-sample-size ones. In addition, we will show in the following sections that sequential estimation is inevitable when the Fisher information observed by sensors is random. We will also provide optimum stopping time analysis for the proposed sequential estimators.

There are a few works considering the sequential decentralized estimation in the literature, e.g., \cite{Papa01,Abdallah01,Zuo11-thesis,Zhao07}, in which sensors employ the conventional uniform-in-time samplers to sample and transmit their local observations. On the other hand, similar to \cite{Fellouris12}, in this paper we will consider using level-triggered sampling, a non-uniform sampling strategy, which perfectly fits to transmitting information in decentralized systems as recently shown in \cite{Fellouris11,Yilmaz11,Yilmaz13}. Level-triggered sampling, eliminating the need for quantization, naturally outputs 1-bit information, which upon transmission produces a high quality recovery at the FC with a very fine resolution (even full resolution if sensors observe continuous-time signals with continuous paths, e.g., Brownian motion). Hence, the level-triggered-sampling-based information transmission, sending 1 bit per sample, enjoys the simplicity of digital transmission, and at the same time it is as powerful as analog transmission producing fine resolution recovery. Furthermore, it provides censoring of unreliable observations, similarly to \cite{Msechu12}.

The decentralized estimators in \cite{Ribeiro06,Ribeiro06-2,Msechu12,Aysal08} involve iterative procedures for solving convex optimization problems. It is concluded in \cite{Ribeiro06} that under relaxed bandwidth constraints the simple-minded quantized sample mean estimator (QSME), in which sensors simply send their quantized observations to the FC, should be preferred over some more complex estimator. Our level-triggered-sampling-based estimators are as simple as QSME, and are designed under strict bandwidth constraints.

In this paper, we use the standard notation to denote the types of convergence of random variables, e.g., $\overset{d}{\rightarrow}$, $\overset{p}{\rightarrow}$, $\overset{a.s.}{\rightarrow}$ and $\overset{L^n}{\rightarrow}$ denote convergence in distribution, convergence in probability, almost sure convergence and convergence in the $n$-th order moment, respectively. Throughout the paper, $\Exp[\cdot]$ and $\text{Var}(\cdot)$ denote expectation and variance, respectively. \textcolor{blue}{We also use the asymptotic notations $o(\cdot)$, $O(\cdot)$, $\Theta(\cdot)$, and $\omega(\cdot)$ in their standard definition \footnote{A quick reminder for the definitions of the notations $o(\cdot)$, $O(\cdot)$, $\Theta(\cdot)$, and $\omega(\cdot)$:
$f(x) = o\left(g(x)\right)$ if $f(x)$ grows with a lower rate than $g(x)$;
$f(x) = O\left(g(x)\right)$ if $f(x)$ grows with a rate that is no larger than the rate of $g(x)$;
$f(x) = \Theta\left(g(x)\right)$ if $f(x)$ grows with exactly the same rate as $g(x)$; and
$f(x) = \omega\left(g(x)\right)$ if $f(x)$ grows with a larger rate than $g(x)$.}.
Particularly, $o(1)$ represents a term that tends to 0, and $O(1)$ represents a constant term.}

The remainder of the paper is organized as follows. We formulate the decentralized estimation problem and provide the necessary background information in Section \ref{sec:prob}. The optimal centralized estimator and decentralized estimators that we propose are described in Section \ref{sec:optCen} and Section \ref{sec:alg}, respectively. In Section \ref{sec:perf}, asymptotic performances of the proposed decentralized estimators are analyzed. Finally, we give simulation results in Section \ref{sec:sim}, and conclude the paper in Section \ref{sec:conc}.

\section{Problem Formulation and Background}
\label{sec:prob}

Consider the problem of \textcolor{blue}{estimating a non-random complex parameter $x$} at a central unit, i.e., the fusion center (FC), via noisy observations collected at $K$ distributed nodes, i.e., sensors. Let $y^k_t,~t\in\bN,~k=1,\ldots,K$, denote the discrete-time noisy sample observed by the $k$-th sensor at time $t$, given by
\be
    \label{eq:observe}
    y^k_t = x h^k_t + w^k_t,
\ee
where $x\in\bC$ is the constant parameter to be estimated, $h^k_t\in\bC$ is the channel gain, random in general, and observed by the $k$-th sensor, and $w^k_t\sim \cN_c(0,\sigma_k^2)$ is the complex Gaussian noise assumed to be independent and identically distributed (i.i.d.) across time and independent but not necessarily identically distributed across sensors. Accordingly, given $h^k_t$ we have $y^k_t \sim \cN_c(x h^k_t,\sigma_k^2)$, i.e., $y_t^k$ is conditionally Gaussian. In the general case  $h_t^k$ is random and assumed to be i.i.d. across time and independent of $w^k_t$, which  corresponds to the fading channels. We will also consider the additive white Gaussian noise (AWGN) channels, where $h_t^k$ is deterministic and $h_t^k=h_k,~\forall k,t$, as a particular case. \textcolor{blue}{Throughout the paper we assume that
\begin{itemize}
\item[] {\bf(A1)}~~$0<|\Re(x)|,|\Im(x)|<\infty$,
\item[] {\bf(A2)}~~$0<|h_t^k|<\infty,~\forall k,t$,
\end{itemize}}
where $\Re(\cdot)$ and $\Im(\cdot)$ denote the real and imaginary parts of a complex number.
Note that these are mild assumptions required for analysis purposes, which do not impose any bounds, but only disregard some impractical cases. Regarding the channel coefficients $\{h_t^k\}$ we only assume (A2), which holds with probability $1$ for all practical AWGN and fading scenarios, hence can be justified almost surely. In other words, we do not assume any specific distribution for $\{h_t^k\}$.
\ignore{In (A1) and (A2), we in fact assume that both the real and imaginary parts in absolute value are greater than zero, which are stated as above for notational simplicity.}

If sensors transmit their observations in whole by using infinite number of bits, then the FC will have access to all local observations $\{y_t^k\}_{t,k}$ \footnote{The subscripts $t$ and $k$ in the set notation denote $t\in\bN$ and $k=1,\ldots,K$, respectively.}, which corresponds to the conventional \emph{centralized} estimation problem. However, in practice, due to power and bandwidth constraints, sensors typically sample their observations and transmit only a few bits per sample to the FC. In such \emph{decentralized} setup, the FC can only obtain a summary of local observations based on which it performs estimation. Obviously, the performance of a decentralized estimator depends on how comprehensive the summary, that the FC receives, is. In other words, the sampling and quantization strategies at sensors, and the fusion rule employed by the FC determine the performance of a decentralized estimator. Since under ideal conditions (i.e., no sampling and infinite-precision quantization) the decentralized estimator becomes the centralized one, the optimal performance of the centralized estimator is a benchmark for decentralized estimators. Hence, we will first analyze the optimal centralized estimator.

Let $\cH_t^k$ denote the set of channel gains observed at the $k$-th sensor up to time $t$, i.e., $\cH_t^k \triangleq \{h_{\tau}^k\}_{\tau}$ \footnote{We use the subscript $\tau$ in the set notation to denote $\tau=1,\ldots,t$.}. Define also $\cH_t \triangleq \{\cH_t^k\}_k$.
In this paper, we are interested in an estimator (centralized or decentralized), $\hat{x}_t$, of $x$, that is conditionally (and unconditionally) unbiased, i.e., $\Exp[\hat{x}_t|\cH_t]=x,~\forall t$, hence $\Exp[\hat{x}_t]=x$, and in minimum time achieves a specified target accuracy in terms of the squared error loss, i.e., $(\hat{x}_{t}-x)^2 \leq 1/\cI$. 
Since $x$ is unknown, we need to estimate the true squared error to assess the accuracy of the estimator. In general, the mean squared error (MSE), $\Exp\left[(\hat{x}_t-x)^2\right]=\text{Var}(\hat{x}_t)$, is used to estimate the true squared error. In this paper, we will use the conditional variance, $\text{Var}\left( \hat{x}_t | \cH_t\right)$, in the presence of an ancillary statistic $\cH_t$ \cite{Efron78}. Note that $\text{Var}(\hat{x}_t)=\Exp\left[ \text{Var}\left( \hat{x}_t | \cH_t\right) \right]$, and whenever $\text{Var}\left( \hat{x}_t | \cH_t\right)$ itself is available, there is no need to use its mean. Hence the conditional variance is a better (in fact the best \cite{Sandved68}) estimate of the true squared error than the unconditional variance.
Thus, we aim to find the conditionally unbiased estimator, $\hat{x}_t$, that satisfies the following inequality,
\be
    \label{eq:obj}
    \text{Var}\big( \hat{x}_{\cT} | \cH_{\cT} \big) \leq \frac{1}{\cI},
\ee
where $\cT$, given in \eqref{eq:stop}, is the minimum time for any conditionally unbiased estimator to achieve the target accuracy $1/\cI$.

The Cramer-Rao lower bound (CRLB), defined using the Fisher information $I_t$, provides the minimum variance for an unbiased estimator of $x$ at time $t$, i.e., $\text{Var}(\hat{x}_t) \geq \text{CRLB} = 1/I_t$ \cite[pp. 171]{Poor94}. Given $\cH_t$, we can define the conditional Fisher information, $I_t^c$, and accordingly the conditional CRLB, $1/I_t^c$, as in \cite{Francos96}. Then, similarly we have $\text{Var}\left( \hat{x}_t | \cH_t\right) \geq 1/I_t^c$. Assuming a conditionally efficient estimator, which achieves $\text{Var}\big( \hat{x}_t | \cH_t \big)=1/I_t^c,\forall t$,  from \eqref{eq:obj} the optimal estimation time (stopping time) $\cT$ is given by
\be
    \label{eq:stop}
    \cT \triangleq \min\{ t \in \bN: I_t^c \geq \cI \},
\ee
as in \cite{Grambsch83}.

Note that the conditional problem has stricter constraints than the unconditional problem.
Conditional unbiasedness implies unconditional unbiasedness, but not vice versa. Moreover, imposing the condition in \eqref{eq:obj} we want to satisfy the target accuracy at each realization, which is a stricter requirement than its unconditional counterpart $\text{Var}( \hat{x}_{\cT}) \leq \frac{1}{\cI}$ aiming to satisfy the target accuracy only on average.
We will next analyze the conditional MLE, which will be shown to be conditionally unbiased and efficient, as the optimal centralized estimator.

\section{Optimal Centralized Estimator}
\label{sec:optCen}

In the centralized setup under fading channels, the $k$-th sensor transmits both $\{y_t^k\}_t$ and $\{h^k_t\}_t$ to the FC by using infinite number of bits, hence both $\{y_t^k\}_{t,k}$ and $\{h_t^k\}_{t,k}$ are available to the FC.
Note that under fading channels $\{y_t^k|h_t^k\}_t$ (across time) are independent, but not identically distributed (i.n.i.d.), and similarly \textcolor{blue}{$\{y_t^k|h_t^k\}_k$ (across sensors) are i.n.i.d.}. Under AWGN channels $\{y_t^k\}_k$ are i.n.i.d. (across sensors), but $\{y_t^k\}_t$ are i.i.d. (across time).

Hence, in general, due to the independence across sensors and time, the conditional log-likelihood $L_t$ of the global observations up to time $t$, $\{y_\tau^k\}_{\tau,k}$, is given by
\be
    L_t = \sum_{k=1}^K L^k_t = \sum_{k=1}^K \sum_{\tau=1}^t l_{\tau}^k \ \ \
    \text{where} \ \ \ l^k_{\tau} \triangleq -\frac{|y^k_{\tau}-x h^k_{\tau}|^2}{\sigma_k^2}-\log \pi \sigma_k^2
\ee
is the conditional log-likelihood of a single observation $y_{\tau}^k$ given $h_{\tau}^k$. The conditional score function for the real part of $x$, $S_t \triangleq \frac{\text{d}}{\text{d}x_r} L_t$, is then written as
\be
    \label{eq:score}
    S_t = \frac{2}{\sigma_k^2}\sum_{k=1}^K \sum_{\tau=1}^t \left[ \Re((h_{\tau}^k)^*y_{\tau}^k)-x_r|h^k_{\tau}|^2\right],
\ee
where $x_r\triangleq\Re(x)$ and $(\cdot)^*$ denotes the complex conjugate of a complex number. Next, we write the conditional observed Fisher information for $x_r$, $U_t \triangleq -\frac{\text{d}}{\text{d}x_r} S_t$, as
\be
    \label{eq:Udef}
    U_t = \sum_{k=1}^K U_t^k = \sum_{k=1}^K \sum_{\tau=1}^t \frac{2|h^k_{\tau}|^2}{\sigma_k^2}.
\ee

The conditional MLE, $\hat{x}_{r,t}$, maximizes $L_t$, hence we have $S_t(\hat{x}_{r,t})=0$.
From (\ref{eq:score}), $\hat{x}_{r,t}$ is then given by
\begin{align}
    \label{eq:MLE}
    \hat{x}_{r,t} =& \frac{\sum_{k=1}^K \sum_{\tau=1}^t \frac{2\Re((h^k_{\tau})^*y^k_{\tau})}{\sigma_k^2}} {\sum_{k=1}^K  \sum_{\tau=1}^t  \frac{2|h^k_{\tau}|^2}{\sigma_k^2}} = \frac{V_t}{U_t} \\
    \label{eq:Vdef}
    \text{where} \ \ \ V_t \triangleq & \sum_{k=1}^K \sum_{\tau=1}^t \frac{2\Re((h^k_{\tau})^*y^k_{\tau})}{\sigma_k^2} = \sum_{k=1}^K V_t^k.
\end{align}
\textcolor{blue}{Similarly, the conditional MLE for the imaginary part of $x$ is written as $\hat{x}_{i,t}=\frac{\bar{V}_t}{U_t}$ where $\bar{V}_t \triangleq \sum_{k=1}^K \sum_{\tau=1}^t \frac{2\Im((h^k_{\tau})^*y^k_{\tau})}{\sigma_k^2}$. Since the estimators for the real and imaginary parts are of the same form, all the discussions and analyses performed in the remainder of the paper hold for both. Therefore, we will henceforth consider only the real part estimator without the subscript.}

We can rewrite (\ref{eq:score}) as $S_t = V_t - x U_t$. Dividing both sides by $U_t$ and using (\ref{eq:MLE}) we get
\be
    \label{eq:MLE2}
    \hat{x}_t = x+\frac{S_t}{U_t}.
\ee
Writing (\ref{eq:Vdef}) explicitly as $V_t = \sum_{k=1}^K \sum_{\tau=1}^t \frac{2(\Re(y^k_{\tau}) \Re(h^k_{\tau}) + \Im(y^k_{\tau}) \Im(h^k_{\tau})) } {\sigma_k^2}$, and noting that $\Re(y^k_{\tau}) \sim \cN(x\Re(h^k_{\tau}), \frac{\sigma_k^2}{2})$, $\Im(y^k_{\tau}) \sim \cN(x\Im(h^k_{\tau}), \frac{\sigma_k^2}{2})$ given $h^k_{\tau}$, we have $V_t \sim \cN(xU_t,U_t)$, and thus $S_t \sim \cN(0,U_t)$ given $\cH_t$. Therefore, from (\ref{eq:MLE2}) we have
\be
    \label{eq:MLEdist}
    \hat{x}_t|\cH_t \sim \cN(x,1/U_t).
\ee

From the definition of the Fisher information, $I_t \triangleq \Exp[S_t^2] = \Exp[U_t]$, we write the conditional Fisher information as
\be
    \label{eq:Finf}
    I_t^c = \Exp\big[ U_t|\cH_t \big] = U_t = \sum_{k=1}^K \sum_{\tau=1}^t \frac{2|h^k_{\tau}|^2}{\sigma_k^2}.
\ee
Hence, we have the following result for the conditional MLE.

\begin{lem}
    \label{lem:cen}
    Assuming (A2) the conditional MLE, $\hat{x}_t$, given in \eqref{eq:MLE}, is conditionally unbiased, i.e., $\Exp\left[\hat{x}_t | \cH_t\right]=x$, consistent, i.e., $\hat{x}_t \overset{p}{\rightarrow} x$ given $\cH_t$ \textcolor{blue}{as $t\to\infty$}, and efficient, i.e., ${\rm Var}\left( \hat{x}_t | \cH_t\right) = 1/I_t^c$, $\forall t$.
\end{lem}

\begin{IEEEproof}
The proof is given in Appendix A.
\end{IEEEproof}

Note that in the particular case of AWGN channels, where we have $h^k_{\tau}=h_k,~\forall \tau$, all results obtained conditional on $\cH_t$ until now, including Lemma \ref{lem:cen}, are valid only in their unconditional forms since $h_k,~\forall k$, is deterministic and known. Hence, in this case the Fisher information, $I_t = \sum_{k=1}^K \frac{2t |h_k|^2}{\sigma_k^2}$, is deterministic. Consequently, the optimal stopping time, $\cT$, defined in (\ref{eq:stop}), is also deterministic and given by
\be
    \label{eq:stoptime_awgn}
    \cT = t_{\cI} = \left\lceil \frac{\cI}{\sum_{k=1}^K \frac{2|h_k|^2}{\sigma_k^2}} \right\rceil
\ee
where $\lceil \cdot \rceil$ is the ceiling operator. Hence, we have the following corollary.

\begin{cor}
\label{cor:awgn}
The fixed-sample-size MLE $\hat{x}_{t_{\cI}}$, which has a variance of $1/I_{t_{\cI}}$ (cf. Lemma \ref{lem:cen}), is the optimal centralized estimator under AWGN channels in terms of the objective in \eqref{eq:obj}.
\end{cor}

Under fading channels, however, the conditional Fisher information $I_t^c$ in (\ref{eq:Finf}), and accordingly the optimal stopping time $\cT$ in (\ref{eq:stop}) are random. Hence in this case, we consider a sequential conditional MLE, $(\cT,\hat{x}_{\cT})$. In \cite[pp. 96]{Ghosh97}, for non-i.i.d. observations, the use of CRLB was extended to sequential estimators. We can further extend it to sequential conditional estimators as stated in the following lemma without proof.

\begin{lem}
\label{lem:seq+con_CRLB}
The conditional variance of a sequential estimator $(\cT,\hat{x}_{\cT})$ that is conditionally unbiased, i.e., $\Exp\left[\hat{x}_{\cT} | \cH_{\cT}\right]=x$, and with a random stopping time $\cT$, is lower bounded by the conditional CRLB, i.e.,
\be
\text{Var}(\hat{x}_{\cT} | \cH_{\cT})\geq\frac{1}{I_{\cT}^c}.
\ee
\end{lem}

Then, we can write the following corollary for the fading case.
\begin{cor}
\label{cor:fade}
The sequential MLE $(\cT,\hat{x}_{\cT})$, having a conditional variance of $1/I_{\cT}^c$, is the optimal centralized estimator under fading channels in terms of the objective in \eqref{eq:obj}.
\end{cor}

\begin{IEEEproof}
    It suffices to show that $\Exp\left[(\hat{x}_{\cT}-x)^2|\cH_{\cT}\right]=1/I_{\cT}^c$. Note that we can write $$\sum_{t=0}^{\infty} \Exp\left[(\hat{x}_t-x)^2 \ind{t=\cT} |\cH_t\right]=\sum_{t=0}^{\infty} \Exp\left[(\hat{x}_t-x)^2 |\cH_t\right]\ind{t=\cT},$$
    where $\ind{\cdot}$ is the indicator function, since $U_t$ depends only on $\cH_t$ and having $I_t^c=U_t$ from \eqref{eq:Finf} the event $\{\cT=t\}$ is deterministic given $\cH_t$ [cf. \eqref{eq:stop}]. From Lemma \ref{lem:cen}, we have $\Exp\left[(\hat{x}_t-x)^2 |\cH_t\right]=1/I_t^c$, hence $$\Exp\left[(\hat{x}_{\cT}-x)^2|\cH_{\cT}\right]=\sum_{t=0}^{\infty} \frac{1}{I_t^c} \ind{t=\cT} = \frac{1}{I_{\cT}^c},$$ which concludes the proof.
\end{IEEEproof}

Note that we were able to obtain the optimal sequential estimator, that achieves the conditional sequential CRLB, since our stopping time $\cT$ depends only on the channels, i.e., $\cH_t$, but not on the observations $\{y_t^k\}$. In general, for a stopping time that also depends on the observations it was shown in \cite{Ghosh87} that the sequential CRLB is not attainable under any distribution except for Bernoulli distribution.
In the following section, following the optimal centralized estimators in Corollary \ref{cor:awgn} and Corollary \ref{cor:fade}, we will propose decentralized estimators based on either the level-triggered sampling or the traditional uniform-in-time sampling. And in Section \ref{sec:perf}, we will analyze the conditions under which the decentralized estimators given in Section \ref{sec:alg} achieve asymptotic unbiasedness, consistency and asymptotic optimality.

\section{Decentralized Estimators}
\label{sec:alg}

In this section, we will develop decentralized estimators, $(\tilde{\cT},\tilde{x}_{\tilde{\cT}})$, by imitating the optimal centralized estimators given in the previous section. We will start with the case of AWGN channels, and then continue with the general case of fading channels.

\subsection{AWGN Channels}
\label{sec:awgn_alg}

Note that the optimal centralized estimator is computed using both $U_t$ and $V_t$ [cf. \eqref{eq:MLE}], whereas the optimal stopping time is determined using only $U_t$ [cf. \eqref{eq:stop} and \eqref{eq:Finf}].
In this case, since we have $h^k_{\tau}=h_k,~\forall \tau$, from \eqref{eq:Udef}, $U_t = \sum_{k=1}^K \frac{2t |h_k|^2}{\sigma_k^2}$ is deterministic, and thus can be known by the FC beforehand. Hence,
the optimal stopping time $\tilde{\cT}$ is deterministic and given by \eqref{eq:stoptime_awgn}. In other words, under AWGN channels the fixed-sample-size decentralized estimator $\tilde{x}_{t_{\cI}}$ is of interest. In a decentralized system, $V_t$ given in (\ref{eq:Vdef}) is a random process unlike $U_t$, and thus is not readily available to the FC. From Corollary \ref{cor:awgn}, and (\ref{eq:MLE}), we see that $V_{t_{\cI}}$ is a sufficient statistic for optimally estimating $x$, hence sensors should report $\{V_{t_{\cI}}^k\}_k$ to the FC.
This can be done either sequentially or once at the optimal stopping time, $t_{\cI}$, using the same number of bits in total on average.

The sequential approach, by its nature, has a number of advantages in practice over the fixed-time approach. Firstly, in the sequential approach, early estimates before the stopping time, i.e., $\{\tilde{x}_t: t<t_{\cI}\}$, are available, although they are not as accurate as the final estimate $\tilde{x}_{t_{\cI}}$. This is a useful feature especially when $t_{\cI}$ is large. Secondly, in the sequential approach, each sensor sends several small messages to the FC until $t_{\cI}$, requiring significantly less bandwidth than sending a single large message at time $t_{\cI}$ in the fixed-time approach. Moreover, in the fixed-time approach there is a possibility of congestion at the FC due to the burst of bits received at time $t_{\cI}$.

In this paper, following the sequential approach we propose a decentralized MLE based on level-triggered sampling, which we call LT-DMLE. Note that LT-DMLE is still a fixed-sample-size estimator despite the fact that it sequentially reports $\{V_{t_{\cI}}^k\}_k$ to the FC. We will describe first the conventional decentralized MLE (DMLE) following the fixed-time approach, and then LT-DMLE.

\subsubsection{DMLE}
\label{sec:dmle}

Each sensor $k$ following the fixed-time approach, at time $t_{\cI}$, quantizes $V_{t_{\cI}}^k$ into $\tilde{V}_{t_{\cI}}^k$ using a traditional mid-riser uniform quantizer with the step size $\frac{t_{\cI}\phi_k}{2^{R_k-1}}$, and transmits $R_k$ bits to the FC. The parameter $\phi_k$ is selected such that $\Pro\left(|V_{t_{\cI}}^k|>t_{\cI}\phi_k\right)$ is sufficiently small so that $\Exp\big[|\tilde{V}_{t_{\cI}}^k-V_{t_{\cI}}^k|\big]<\frac{t_{\cI}\phi_k}{2^{R_k}}$.
Specifically, the interval $[-t_{\cI}\phi_k,t_{\cI}\phi_k]$ is uniformly partitioned into $2^{R_k}$ subintervals, and for each subinterval its mid value is used as the quantization level. The $V_{t_{\cI}}^k$ values that are out of the interval $[-t_{\cI}\phi_k,t_{\cI}\phi_k]$ are mapped to the closest quantization level, i.e., the values satisfying $V_{t_{\cI}}^k>t_{\cI}\phi_k$ and $V_{t_{\cI}}^k<-t_{\cI}\phi_k$ are quantized as $\tilde{V}_{t_{\cI}}^k=t_{\cI}\phi_k\frac{2^{R_k}-1}{2^{R_k}}$ and $\tilde{V}_{t_{\cI}}^k=-t_{\cI}\phi_k\frac{2^{R_k}-1}{2^{R_k}}$, respectively.

The FC, upon receiving $R_k$ bits from each sensor at time $t_{\cI}$, recovers $\tilde{V}_{t_{\cI}}^k,~k=1,\ldots,K$, and then computes
\be
    \label{eq:dmle1}
    \tilde{V}_{t_{\cI}}=\sum_{k=1}^K \tilde{V}_{t_{\cI}}^k.
\ee
Finally, similar to (\ref{eq:MLE}) the estimate
\be
    \label{eq:dmle2}
    \tilde{x}_{t_{\cI}}=\frac{\tilde{V}_{t_{\cI}}}{U_{t_{\cI}}}
\ee
is formed \footnote{DMLE corresponds to the quantized sample mean estimator (QSME) in \cite{Ribeiro06}.}.

\subsubsection{LT-DMLE}
\label{sec:lt-dmle}

For LT-DMLE, following the sequential approach, we propose that each sensor $k$, via level-triggered sampling, informs the FC whenever considerable change occurs in its local process $V_t^k$. The level-triggered sampling is a simple form of event-triggered sampling, in which sampling (communication) times $\{t_{n,V}^k\}_n$ \footnote{The subscript $n$ in the set notation denotes $n\in\bN$.} are not deterministic, but rather dynamically determined by the random process $V_t^k$, i.e.,
\be
    \label{eq:samp}
    t_{n,V}^k \triangleq \min\{t>t_{n-1,V}^k : V_t^k-V_{t_{n-1,V}^k}^k \not\in (-d_k,d_k)\},~n\in\bN,~t_{0,V}^k=0.
\ee
The threshold parameter $d_k$ is a constant known by both sensor $k$ and the FC.

At each sampling time $t_{n,V}^k$, sensor $k$ transmits $r_V$ bits, $b_{n,1}^k b_{n,2}^k \ldots b_{n,r_V}^k$, to the FC. The first bit, $b_{n,1}^k$, indicates the threshold crossed (either $d_k$ or $-d_k$) by the incremental process $v_n^k \triangleq V_{t_{n,V}^k}^k-V_{t_{n-1,V}^k}^k$, i.e.,
\be
    \label{eq:sign}
    b_{n,1}^k=\text{sign}(v_n^k).
\ee
The remaining $(r_V-1)$ bits, $b_{n,2}^k \ldots b_{n,r_V}^k$, are used to quantize the over(under)shoot $q_n^k \triangleq |v_n^k|-d_k$ into $\tilde{q}_n^k$. At each sampling time $t_{n,V}^k$, the overshoot value $q_n^k$ cannot exceed the magnitude of the last sample $\frac{2}{\sigma_k^2}\big|\Re((y^k_{t_n^k})^*h^k_{t_n^k})\big|$ in the incremental process $v_n^k=\sum_{\tau=t_{n-1,V}^k+1}^{t_{n,V}^k} \frac{2\Re((y^k_{\tau})^*h^k_{\tau})}{\sigma_k^2}$. Hence, the interval $[0,\phi_k]$ is uniformly divided into $2^{r_V-1}$ subintervals using again a mid-riser quantizer with the step size $\frac{\phi_k}{2^{r_V-1}}$. Here the parameter $\phi_k$ is determined such that $\Pro(q_n^k>\phi_k)$ is sufficiently small so that $\Exp\big[|\tilde{q}_n^k-q_n^k|\big]<\frac{\phi_k}{2^{r_V}}$.

Note that if $V_t^k$ were a continuous-time process with continuous paths, e.g., Brownian motion, then it would exactly hit the thresholds, i.e., no overshoot would occur, and thus no quantization bits would be needed, i.e., $r_V=1$. The threshold parameter $d_k$ is determined so that the $k$-th sensor, up to time $t_{\cI}$, transmits on average $R_k$ bits to the FC, i.e., communicates to the FC on average $\frac{R_k}{r_V}$ times.

The FC, upon receiving the bits $b_{n,1}^k b_{n,2}^k \ldots b_{n,r_V}^k$ from the sensor $k$ at time $t_{n,V}^k$, recovers the quantized value of $v_n^k$ by computing
\be
    \label{eq:VincrRecov}
    \tilde{v}_n^k \triangleq b_{n,1}^k(d_k+\tilde{q}_n^k).
\ee
Then, it sequentially sums up $\{\tilde{v}_n^k\}_{n,k}$, at the sampling (communication) times $\{t_{n,V}^k\}_{n,k}$ to obtain an approximation $\tilde{V}_t$ to the sufficient statistic $V_t$, i.e.,
\be
    \label{eq:VRecov}
    \tilde{V}_t \triangleq \sum_{k=1}^K \sum_{n=1}^{N_t^k} \tilde{v}_n^k = \sum_{k=1}^K \tilde{V}_t^k,
\ee
where $N_t^k$ is the number of messages that the FC receives from the sensor $k$ about $V_t^k$ up to time $t$.
During the times the FC receives no message, i.e., $t \not\in \{t_{n,V}^k\}_{n,k}$, $\tilde{V}_t$ is kept constant.
Replacing $V_t$ with $\tilde{V}_t$ in (\ref{eq:MLE}) the following decentralized estimator,
\be
    \label{eq:decEst}
    \tilde{x}_t = \frac{\tilde{V}_t}{U_t},
\ee
is obtained at the FC. Finally, the scheme stops at time $\tilde{\cT}=t_{\cI}$ [cf. \eqref{eq:stoptime_awgn}] after computing the final estimate $\tilde{x}_{t_{\cI}}=\frac{\tilde{V}_{t_{\cI}}}{U_{t_{\cI}}}$.

\subsection{Fading Channels}
\label{sec:fade_alg}

Under fading channels, $U_t$ is random, hence sensors should report both $\{V_t^k\}_k$ and $\{U_t^k\}_k$ to the FC. In this case, only the sequential approach can be used to report $\{U_t^k\}_k$ to the FC since the stopping (optimal estimation) time, $\cT$, is random. A straightforward way to sequentially report $\{U_t^k\}_k$ is to use a conventional uniform-in-time sampler followed by a quantizer. Alternatively, level-triggered sampling can be employed, which has certain advantages over the uniform-in-time sampling, as will be shown in Section \ref{sec:perf}.
On the other hand, $\{V_t^k\}_k$, as in the AWGN case, can be reported to the FC either sequentially or once at time $\cT$, when the process stops.
Hence, we propose two sequential decentralized MLEs based on level-triggered sampling, and two based on uniform sampling. In the first group of estimators, $\{U_t^k\}_k$ are sequentially reported, but $\{V_t^k\}_k$ are reported once at time $\cT$, hence the names LT-sDMLE (level-triggered sampling based sequential DMLE) and U-sDMLE (uniform sampling based sequential DMLE) are used. In the second group, both $\{U_t^k\}_k$ and $\{V_t^k\}_k$ are sequentially reported, hence we name the estimators LT-dsDMLE (level-triggered sampling based doubly sequential DMLE) and U-dsDMLE (uniform sampling based doubly sequential DMLE). We next explain these four estimators in detail.

\subsubsection{LT-sDMLE}
\label{sec:lt-sdmle}

In LT-sDMLE, sensors sample only $\{U_t^k\}_k$ via level-triggered sampling at the following sampling times,
\be
    \label{eq:sampU}
    t_{n,U}^k \triangleq \min\{t>t_{n-1,U}^k : U_t^k-U_{t_{n-1,U}^k}^k \geq e_k\},~n\in\bN,~t_{0,U}^k=0,
\ee
where the threshold $e_k$ is a constant chosen by the designer and made available to the FC and sensor $k$. Note that in \eqref{eq:sampU} we use a single threshold different from \eqref{eq:samp} since $U_t^k$, given in \eqref{eq:Udef}, is a nondecreasing process. Define the incremental process $u_n^k \triangleq U_{t_{n,U}^k}^k-U_{t_{n-1,U}^k}^k$.
At each sampling time $t_{n,U}^k$, sensor $k$ transmits $r_U$ bits to the FC, all of which are used to quantize the overshoot $p_n^k \triangleq u_n^k-e_k$ into $\tilde{p}_n^k$ using a mid-riser uniform quantizer with the step size $\frac{\theta_k}{2^{r_U}}$, similar to LT-DMLE. In this case, we do not need to allocate a sign bit. The overshoot $p_n^k$ is bounded by the last sample $\frac{2 \big|h^k_{t_{n,U}^k}\big|^2}{\sigma_k^2}$ in the incremental process $u_n^k=\sum_{\tau=t_{n-1,U}^k+1}^{t_{n,U}^k} \frac{2 |h^k_{\tau}|^2}{\sigma_k^2}$.
Specifically, each sensor $k$ uniformly partitions the interval $[0,\theta_k]$ into $2^{r_U}$ subintervals, where $\theta_k$ is selected such that $\Pro(p_n^k>\theta_k)$ is sufficiently small so that $\Exp\big[|\tilde{p}_n^k-p_n^k|\big]<\frac{\theta_k}{2^{r_U+1}}$; and then at each sampling time $t_{n,U}^k$ determines the quantization level for $p_n^k$ and transmits its index to the FC using $r_U$ bits.
When the scheme is terminated by the FC at the random stopping time $\tilde{\cT}$, each sensor $k$, as in DMLE, by using a mid-riser uniform quantizer with step size $\frac{\tilde{\cT}\phi_k}{2^{R_k-1}}$ and $R_k$ bits quantizes $V_{\tilde{\cT}}^k$ into $\tilde{V}_{\tilde{\cT}}^k$, which is then transmitted to the FC. The parameter $\phi_k$ is selected such that $\Pro(|V_{\tilde{\cT}}^k|>\tilde{\cT}\phi_k)$ is sufficiently small so that $\Exp\big[|\tilde{V}_{\tilde{\cT}}^k-V_{\tilde{\cT}}^k|~|~\cH_{\tilde{\cT}}\big]<\frac{\tilde{\cT}\phi_k}{2^{R_k}}$.

The FC, upon receiving the $r_U$ bits at time $t_{n,U}^k$, similar to \eqref{eq:VincrRecov} computes
\be
    \label{eq:UincrRecov}
    \tilde{u}_n^k \triangleq e_k+\tilde{p}_n^k.
\ee
Then, similar to \eqref{eq:VRecov} it also computes
\be
    \label{eq:URecov}
    \tilde{U}_t \triangleq \sum_{k=1}^K \sum_{n=1}^{M_t^k} \tilde{u}_n^k = \sum_{k=1}^K \tilde{U}_t^k,
\ee
where $M_t^k$ is the number of messages that the FC receives from sensor $k$ about $U_t^k$ up to time $t$.
The scheme is terminated at the stopping time, $\tilde{\cT}$ [cf. (\ref{eq:stop}), (\ref{eq:Finf})], given by
\be
    \label{eq:Ustop}
    \tilde{\cT} = \min\{ t\in\bN : \tilde{U}_t \geq \cI \}.
\ee
Finally, the FC, as in DMLE, upon receiving $R_k$ bits from each sensor at time $\tilde{\cT}$, recovers $\tilde{V}_{\tilde{\cT}}^k,~\forall k$, and computes $\tilde{V}_{\tilde{\cT}}=\sum_{k=1}^K \tilde{V}_{\tilde{\cT}}^k$, as well as the estimate $\tilde{x}_{\tilde{\cT}} = \frac{\tilde{V}_{\tilde{\cT}}} {\tilde{U}_{\tilde{\cT}}}$.

\subsubsection{LT-dsDMLE}
\label{sec:lt-dsdmle}

In LT-dsDMLE, there are two different sets of sampling times, namely $\{t_{n,U}^k\}_{n,k}$ and $\{t_{n,V}^k\}_{n,k}$. Each sensor $k$, as in LT-sDMLE, at time $t_{n,U}^k$ [cf. (\ref{eq:sampU})] quantizes $p_n^k$ into $\tilde{p}_n^k$, and transmits $r_U$ bits to the FC until the stopping time $\cT$. Similarly, each sensor $k$, as in LT-DMLE, at time $t_{n,V}^k$ [cf. (\ref{eq:samp})] quantizes $v_n^k$ into $\tilde{v}_n^k$, and transmits $r_V$ bits to the FC until the stopping time $\cT$. The quantization parameter $\phi_k$ is selected such that $\Pro(q_n^k>\phi_k)$ is sufficiently small so that $\Exp\big[|\tilde{q}_n^k-q_n^k|~|~\cH_{t_{n,V}^k}^k\big]<\frac{\phi_k}{2^{r_V}}$.

The FC computes $\tilde{u}_n^k$ at time $t_{n,U}^k$ as in \eqref{eq:UincrRecov}, and $\tilde{v}_n^k$ at time $t_{n,V}^k$ as in \eqref{eq:VincrRecov}. Then, it obtains $\tilde{U}_t$ and $\tilde{V}_t$ as in \eqref{eq:URecov} and \eqref{eq:VRecov}, respectively. Next, similar to (\ref{eq:decEst}) the following estimator,
\be
    \label{eq:dec_est}
    \tilde{x}_t = \frac{\tilde{V}_t}{\tilde{U}_t},
\ee
is formed. Finally, the FC terminates the process at time $\tilde{\cT}$, given by \eqref{eq:Ustop},
immediately after the final estimate $\tilde{x}_{\tilde{\cT}} = \frac{\tilde{V}_{\tilde{\cT}}} {\tilde{U}_{\tilde{\cT}}}$ is computed.

\subsubsection{U-sDMLE}
\label{sec:u-sdmle}

In U-sDMLE, each sensor $k$ uniformly samples $\{U_t^k\}_k$ with period $T_U$, i.e., at times $\{m T_U\}_{m\in\bN}$. Specifically, it computes the incremental process $u_{mT_U}^k \triangleq U_{mT_U}^k-U_{(m-1)T_U}^k$ at time $mT_U$. Using a mid-riser quantizer with the step size $\frac{T_U\theta_k}{2^{r_U}}$ it uniformly divides the interval $[0,T_U\theta_k]$ into $2^{r_U}$ subintervals. Then, at time $mT_U$, it quantizes $u_{mT_U}^k$ into $\tilde{u}_{mT_U}^k$, and transmits the corresponding quantization level index to the FC using $r_U$ bits. Here $\theta_k$ is selected such that $\Pro(u_{mT_U}^k>T_U\theta_k)$ is sufficiently small so that $\Exp\big[|\tilde{u}_{mT_U}^k-u_{mT_U}^k|\big]<\frac{T_U\theta_k}{2^{r_U+1}}$.
When the process stops at time $\tilde{\cT}$, each sensor $k$, as in LT-sDMLE, quantizes $V_{\tilde{\cT}}^k$ into $\tilde{V}_{\tilde{\cT}}^k$ using $R_k$ bits, and then transmits the quantization bits to the FC.

The FC, at time $mT_U$, computes $\tilde{u}_{mT_U}^k$ using the received $r_U$ bits. Then, similar to \eqref{eq:URecov}, it computes
\be
    \label{eq:uni_recov}
    \tilde{U}_t \triangleq \sum_{k=1}^K \sum_{m=1}^{M_t} \tilde{u}_{mT_U}^k,
\ee
where $M_t=\lfloor t/T_U \rfloor$ is the number of sampling (communication) times, until time $t$, for $\{U_t^k\}_k$, and $\lfloor\cdot\rfloor$ is the floor operator. At time $\tilde{\cT}$, given in \eqref{eq:Ustop}, the FC, as in DMLE and LT-sDMLE, terminates the process; recovers $\tilde{V}_{\tilde{\cT}}^k$ upon receiving $R_k$ bits; and finally computes $\tilde{V}_{\tilde{\cT}}$ and the estimate $\tilde{x}_{\tilde{\cT}} = \frac{\tilde{V}_{\tilde{\cT}}} {\tilde{U}_{\tilde{\cT}}}$.

\subsubsection{U-dsDMLE}
\label{sec:u-dsdmle}

We also have two sets of sampling times in U-dsDMLE, for $\{U_t^k\}_k$ and $\{V_t^k\}_k$, that are uniform in time with periods $T_U$ and $T_V$, respectively, i.e., $\{m T_U\}_m$ and $\{m T_V\}_m$. At time $mT_U$, as in U-sDMLE, each sensor $k$ computes $u_{mT_U}^k$; quantizes it into $\tilde{u}_{mT_U}^k$ using $r_U$ bits; and transmits the quantization bits to the FC. Similarly, at time $mT_V$, each sensor $k$ computes the incremental process $v_{mT_V}^k \triangleq V_{mT_V}^k-V_{(m-1)T_V}^k$ and quantizes it into $\tilde{v}_{mT_V}^k$ using a mid-riser quantizer with the step size $\frac{T_V\phi_k}{2^{r_V-1}}$. In particular, the interval $(-T_V\phi_k,T_V\phi_k)$ is uniformly divided into $2^{r_V}$ subintervals, where $\phi_k$ is determined such that $\Pro(|v_{mT_V}^k|>T_V\phi_k)$ is sufficiently small so that $\Exp\big[|\tilde{v}_{mT_V}^k-v_{mT_V}^k|~|~\cH_{mT_V}^k\big]<\frac{T_V\phi_k}{2^{r_V}}$.
Finally, each sensor $k$ at each sampling time $mT_V$ transmits the index of the quantization level corresponding to $\tilde{v}_{mT_V}^k$ to the FC using $r_V$ bits.

The FC, as in U-sDMLE, computes $\tilde{u}_{mT_U}^k$ at time $mT_U$, and also $\tilde{U}_t$ given by \eqref{eq:uni_recov}. Similarly, at time $mT_V$, it computes $\tilde{v}_{mT_V}^k$ using the received $r_V$ bits. Next, similar to \eqref{eq:VRecov}, it computes
\be
    \label{eq:uni_recov_V}
    \tilde{V}_t \triangleq \sum_{k=1}^K \sum_{m=1}^{N_t} \tilde{v}_{mT_V}^k,
\ee
where $N_t=\lfloor t/T_V \rfloor$ is the numbers of sampling times, until time $t$, for $\{V_t^k\}_k$. Using the approximations in \eqref{eq:uni_recov} and \eqref{eq:uni_recov_V}, the estimator $\tilde{x}_t$ is computed as in \eqref{eq:dec_est}, at time $t$. The stopping time of the scheme is given by \eqref{eq:Ustop}.

\section{Performance Analysis}
\label{sec:perf}

In this section, we will derive the conditions under which the decentralized estimators outlined in the previous section are, \textcolor{blue}{as $\cI\to\infty$} and given $\cH_{\tilde{\cT}}$, asymptotically unbiased, i.e., $\Exp[\tilde{x}_{\tilde{\cT}}|\cH_{\tilde{\cT}}] \to x$, consistent, i.e., $\tilde{x}_{\tilde{\cT}} \overset{p}{\rightarrow} x$, and asymptotically optimal. An estimator $\tilde{x}_t$ is said to be asymptotically optimal if $\sqrt{I_t} (\tilde{x}_t-x)$ converges in distribution to a standard Gaussian random variable, i.e., $\sqrt{I_t} (\tilde{x}_t-x) \overset{d}{\rightarrow} \cN(0,1)$, as $t \to \infty$ \cite[pp. 185]{Poor94}.
In our case, we let the target Fisher information $\cI\to\infty$, thus for asymptotic optimality we need to show that
\be
\sqrt{I_{\tilde{\cT}}^c} (\tilde{x}_{\tilde{\cT}}-x) \overset{d}{\rightarrow} \cN(0,1),
\ee
given $\cH_{\tilde{\cT}}$. Note that asymptotic optimality, which is related to the probability distribution, does not imply asymptotic efficiency, i.e., $\Exp[(\tilde{x}_{\tilde{\cT}}-x)^2|\cH_{\tilde{\cT}}]\to1/I_{\tilde{\cT}}^c$, which is related to the second moment.

\subsection{AWGN Channels}
\label{sec:perf_awgn}

The following theorem gives the conditions under which DMLE, following the fixed-time approach, is asymptotically unbiased, consistent, and asymptotically optimal.

\begin{thm}
    \label{thm:awgn1}
    \textcolor{blue}{Assuming (A2)} the decentralized estimator DMLE, given in Section \ref{sec:awgn_alg} is, as $\cI\to\infty$, asymptotically unbiased, i.e., $\Exp[\tilde{x}_{t_{\cI}}-x]\to0$, and consistent, i.e., $\tilde{x}_{t_{\cI}} \overset{p}{\rightarrow} x$, if $R_k\to\infty$ at any rate, $\forall k$. It is also asymptotically optimal, i.e., $\sqrt{I_{t_{\cI}}} (\tilde{x}_{t_{\cI}}-x) \overset{d}{\rightarrow} \cN(0,1)$, if $R_k\to\infty$ at a faster rate than $\log \cI$, i.e., $R_k=\omega(\log \cI),~\forall k$.
\end{thm}

\begin{IEEEproof}
The proof can be found in Appendix B.
\end{IEEEproof}

Now, we proceed to analyze LT-DMLE, that follows the sequential approach to report $\{V_t^k\}_k$, but is still a fixed-sample-size estimator. The next two theorems give the conditions for LT-DMLE to be asymptotically unbiased, consistent, and asymptotically optimal.

\begin{thm}
    \label{thm:awgn3}
    Consider the decentralized estimator LT-DMLE, given in Section \ref{sec:awgn_alg}.
    It is, as $\cI\to\infty$ and \textcolor{blue}{under (A2)}, asymptotically unbiased, i.e., $\Exp[\tilde{x}_{t_{\cI}}-x]\to0$, and consistent, i.e., $\tilde{x}_{t_{\cI}} \overset{p}{\rightarrow} x$, if $d_k\to\infty$ at a slower rate than $\cI$, i.e., $d_k=o(\cI),~\forall k$.
\end{thm}

\begin{IEEEproof}
    The proof is presented in Appendix C.
\end{IEEEproof}

\ignore{
\underline{\it{Remark:}}
    \bi
    \item Since sensors observe discrete-time signals, over(under)shooting the sampling thresholds causes an inevitable discrepancy between the real value of the local process $V_t^k$ and the recovered value $\tilde{V}_t^k$ at the FC. This discrepancy propagates with the arrival of each new message as time passes, causing performance loss. Letting the sampling threshold $d_k$ tend to infinity as $\cI\to\infty$ assures consistency and asymptotic unbiasedness since it decreases the communication frequency, as a result of which the overshoot accumulation problem is less emphasized. On the other hand, decreasing communication frequency, i.e., $d_k\to\infty$, is a practically desired feature as it corresponds to consuming less bandwidth.
    \item It is sufficient for consistency and asymptotic unbiasedness to have $r_V\to\infty$ at an arbitrarily slow rate.
    \ei
}

\begin{thm}
    \label{thm:awgn4}
    \textcolor{blue}{Assuming (A1) and (A2)} the decentralized estimator LT-DMLE, given in Section \ref{sec:awgn_alg}, is, as $\cI\to\infty$, asymptotically optimal, i.e., $\sqrt{I_{t_{\cI}}} (\tilde{x}_{t_{\cI}}-x) \overset{d}{\rightarrow} \cN(0,1)$, if $d_k=o(\sqrt{\cI})$ and $r_V=\omega(\log(\sqrt{\cI}/d_k)),~\forall k$.
\end{thm}

\begin{IEEEproof}
    The proof is provided in Appendix D.
\end{IEEEproof}

Note that there are two sources of discrepancy in the sequential estimator based on level-triggered sampling, LT-DMLE. One source is the discrepancy in the messages, i.e., overshoot quantization error, represented by the first terms inside the parentheses in \eqref{eq:thm_aw3_2} and \eqref{eq:thm_aw4_1}. The other source is the missing statistics at the FC, between the last sampling times of the sensors and the stopping time, represented by the second terms inside the parentheses in \eqref{eq:thm_aw3_2} and \eqref{eq:thm_aw4_1}. Having the sampling threshold, $d_k\to\infty$, as $\cI\to\infty$, de-emphasizes the first source since the number of messages decreases, and so does the accumulation of the overshoot quantization error. However, having $d_k\to\infty$ emphasizes the second source since the sampling intervals increase, and so do the missing statistics within the incomplete sampling intervals. Therefore, while having $d_k\to\infty,~\forall k,$ as fast as possible is practically desired since it corresponds to asymptotically low communication rates, there is a trade-off in determining its rate as can be seen in \eqref{eq:thm_aw3_2} and \eqref{eq:thm_aw4_1}. Its rate is upper bounded by $\cI$, and $\sqrt{\cI}$ for asymptotic unbiasedness/consistency (Thm. \ref{thm:awgn3}), and asymptotic optimality (Thm. \ref{thm:awgn4}), respectively.

On the other hand, we want the number of bits, $r_V$, to be as small as possible since it corresponds to low bandwidth usage. To ensure asymptotic unbiasedness/consistency we can keep $r_V$ constant, whereas to ensure asymptotic optimality there is a lower bound, $\log(\sqrt{\cI}/d_k)$, on its rate (Thm. \ref{thm:awgn4}). However, note that having the rate of $d_k$ arbitrarily close to $\sqrt{\cI}$, which is the most practically desired choice for asymptotic optimality, we can have the rate of $r_V$ arbitrarily slow. The bandwidth usage in DMLE is determined by $R_k$ since it is the number of bits transmitted at time $t_{\cI}$.

It is well known that the energy consumption in sensors is mostly due to the data transmission. Hence, the number of transmitted bits $R_k$ by sensor $k$ in DMLE should be as small as possible to meet the energy constraint. However, in Theorem \ref{thm:awgn1} it is lower-bounded by the conditions $R_k\to\infty$ at any rate and $R_k=\omega(\log\cI)$ for asymptotic unbiasedness/consistency and asymptotic optimality, respectively. Note from Section \ref{sec:lt-dmle} that $R_k$ is also the average number of transmitted bits by sensor $k$ until the stopping time in LT-DMLE, i.e., $R_k=\Exp[N_{t_{\cI}}^k]~r_V$. Since the transmitted messages, $\{\tilde{v}_n^k\}_n$, are i.i.d., $N_t^k$ is a renewal process. By the elementary renewal theorem we have $\frac{\Exp[N_{t_{\cI}}^k]}{t_{\cI}}\to\frac{1}{\Exp[t_{1,V}^k]}$ as $t_{\cI}\to\infty$, i.e., as $\cI\to\infty$, where it is known that $t_{\cI}=\Theta(\cI)$ from \eqref{eq:stoptime_awgn} and (A2); and $\Exp[t_{1,V}^k]=\Theta(d_k)$ from the proof of Theorem \ref{thm:awgn4}. Therefore, for LT-DMLE we write $R_k=\Theta\Big(\frac{\cI r_V}{d_k}\Big)$ as $\cI\to\infty$. Using Theorem \ref{thm:awgn3} and Theorem \ref{thm:awgn4} we obtain the conditions on $R_k$ in LT-DMLE as $R_k\to\infty$ at any rate and $R_k=\omega(\sqrt{\cI})$ for asymptotic unbiasedness/consistency and asymptotic optimality, respectively.

\begin{table}[t]
  \centering
  \begin{tabular}{|p{2cm}|p{3.5cm}|p{3.5cm}|}
  \hline
  & \centering{Energy} & \hspace{1cm}Bandwidth \\
  \hline
  DMLE \newline (Thm. \ref{thm:awgn1})& AU\&C:~~$E\cong O(1)$ \newline AO:~~$E\cong \Theta(\log \cI)$ & AU\&C:~~$\bar{B}\cong O(1)$ \newline AO:~~$\bar{B}\cong \Theta(\log \cI)$ \\
  \hline
  LT-DMLE \newline (Thm. \ref{thm:awgn3}~\&~\ref{thm:awgn4})& AU\&C:~~$E\cong O(1)$ \newline AO:~~$E\cong \Theta(\sqrt{\cI})$ & AU\&C:~~$B=O(1)$ \newline AO:~~$B\cong O(1)$ \\
  \hline
  \multicolumn{3}{|c|}{AU\&C:~asymptotic unbiasedness and consistency,~~AO:~asymptotic optimality} \\
  \hline
  \end{tabular}
  \caption{Practical implications of Theorems \ref{thm:awgn1}--\ref{thm:awgn4}.}
  \label{tab:awgn}
\end{table}

\textcolor{blue}{We showed that for DMLE $R_k$ represents both the energy and bandwidth consumptions of sensor $k$, whereas for LT-DMLE $R_k$ and $r_V$ represent the energy and bandwidth consumptions, respectively. In Table \ref{tab:awgn}, using Theorems \ref{thm:awgn1}, \ref{thm:awgn3} and \ref{thm:awgn4} we present the slowest (achievable or almost achievable) growth rates for energy and bandwidth usages to ensure asymptotic unbiasedness/consistency and asymptotic optimality. The symbols $=$ and $\cong$ are used to denote achievable and almost achievable rates, respectively. We also use the bar symbol to express the fact that the bandwidth usage in DMLE is already (non-asymptotically) high compared to that in LT-DMLE. It is clearly seen in Table \ref{tab:awgn} that in terms of bandwidth usage the sequential approach, i.e., LT-DMLE, has a big advantage over the fixed-time approach, i.e., DMLE. As the target accuracy level gets finer, i.e., $\cI\to\infty$, LT-DMLE using level-triggered sampling with a low constant bandwidth usage, i.e., constant and small $r_V$, achieves asymptotic unbiasedness and consistency. Furthermore, it can achieve asymptotic optimality by increasing its low bandwidth usage at an arbitrarily slow rate. Whereas, the bandwidth usage of the conventional estimator DMLE, which is even non-asymptotically high, needs to grow slowly and as fast as $\log\cI$ for asymptotic unbiasedness/consistency and asymptotic optimality, respectively.
The growth rates in energy usage are similar except for asymptotic optimality, where DMLE consumes asymptotically less energy than LT-DMLE \big($\log\cI$ vs. $\sqrt{\cI}$\big). We should note also that LT-DMLE has a number of advantages in practice over DMLE, as discussed in Section \ref{sec:awgn_alg}.}

\ignore{
To be able to compare DMLE with LT-DMLE, let us analyze the conditions on $R_k$ in Theorem \ref{thm:awgn1}, shown in the first row of Table \ref{tab:awgn}. By definition, $R_k$ is the average number of bits transmitted until the stopping time by sensor $k$, i.e., $R_k=\Exp[N_{t_{\cI}}^k]~r_V$. Note that $N_{t_{\cI}}^k$ is a renewal process since the received messages, $\{\tilde{v}_n^k\}_n$, are i.i.d., hence using Wald's identity we can write $\Exp[N_{t_{\cI}}^k]=t_{\cI}/\Exp[t_{1,V}^k]$, where it is known that $t_{\cI}=\Theta(\cI)$ and $\Exp[t_{1,V}^k]=\Theta(d_k)$ from \eqref{eq:stoptime_awgn} and the proof of Theorem \ref{thm:awgn4}, respectively. Therefore, paraphrasing the first part of Theorem \ref{thm:awgn1}, we can say that DMLE is asymptotically unbiased and consistent if $r_V=O(1)$, and $d_k\to\infty$ such that $d_k=o(\cI)$, exactly the same set of conditions required for LT-DMLE (cf. Table \ref{tab:awgn}). Similarly, we can rephrase the condition in the second part of Theorem \ref{thm:awgn1} as $r_V=O(1)$, and $d_k\to\infty$ such that $d_k=o(\cI/\log\cI)$. Note that for LT-DMLE, the condition on $r_V$ to achieve asymptotic optimality is similar to the one here for DMLE as $r_V$ can be made arbitrarily close to $O(1)$ for LT-DMLE. However, for DMLE, the condition on $d_k$ to achieve asymptotic optimality is more favorable than the one for LT-DMLE \big[$o(\cI/\log\cI)$ vs. $o(\sqrt{\cI})$\big]. In other words, in DMLE the communication overhead can be asymptotically lower than that in LT-DMLE. On the other hand, LT-DMLE has a number of advantages in practice over DMLE, as discussed in Section \ref{sec:awgn_alg}.
}

\subsection{Fading Channels}
\label{sec:perf_fade}

Under fading channels, $\{y_t^k|h_t^k\}_{t,k}$ are independent, but not identically distributed across sensors and time. Hence, in our derivations for the counting processes such as $t_{1,V}^k$ and $N_{\tilde{\cT}}^k$, and the related processes we cannot use the regular renewal theory identities, including Wald's identity. Another challenge in this case is that the stopping time, $\tilde{\cT}$, is random. In this section, we will first analyze the estimators LT-sDMLE and LT-dsDMLE based on level-triggered sampling, in the first four theorems, and then the estimators U-sDMLE and U-dsDMLE based on uniform sampling, in the last two theorems. Before proceeding to the theorems, we present a number of technical lemmas. From now on, $\bExp[\cdot]$ will denote the conditional expectation given $\cH_t^k$, e.g., $\bExp[V_t^k]=\Exp[V_t^k|\cH_t^k]$, or $\cH_t$, e.g., $\bExp[V_t]=\Exp[V_t|\cH_t]$.

\begin{lem}
    \label{lem:appA}
    \textcolor{blue}{For LT-sDMLE and LT-dsDMLE, the stopping time, $\tilde{\cT}$, under (A2) as $\cI\to\infty$, tends to infinity at the same rate as $\cI$, i.e., $\tilde{\cT}=\Theta(\cI)$ if $e_k$ either remains constant or tends to infinity such that $e_k=O(\cI),~\forall k$. And for U-sDMLE and U-dsDMLE, $\tilde{\cT}=\Theta(\cI)$ if $T_U=O(\cI)$.}
\end{lem}

\begin{IEEEproof}
The proof is given in Appendix E.
\end{IEEEproof}

Note that in the above lemma the condition for LT-sDMLE and LT-dsDMLE reads $e_k=O(\cI)$ except $e_k\to0$ for some $k$. The excluded condition is not of practical importance since it is practically desired that $e_k\to\infty$ or $e_k=O(1)$ for low communication rates. Hence, for practical purposes we can rephrase the condition as $e_k=O(\cI)$.
Let us now analyze, in the following lemma, the asymptotic growth rate of the discrepancy between the global process $U_t$, and its approximation $\tilde{U}_t$.

\begin{lem}
    \label{lem:appB}
\textcolor{blue}{    For LT-sDMLE and LT-dsDMLE with $r_U=O(1)$, under (A2) as $\cI\to\infty$, $|U_{\tilde{\cT}}-\tilde{U}_{\tilde{\cT}}|=o(\cI)$ if $e_k\to\infty$ such that $e_k=o(\cI),~\forall k$. And for U-sDMLE and U-dsDMLE, $|U_{\tilde{\cT}}-\tilde{U}_{\tilde{\cT}}|=o(\cI)$ if $r_U\to\infty$ at any rate.}
\end{lem}

\begin{IEEEproof}
The proof can be found in Appendix F.
\end{IEEEproof}

In the last lemma, we will analyze the asymptotic growth rate of the expected conditional score function in absolute value.

\begin{lem}
    \label{lem:appC}
    For LT-sDMLE and LT-dsDMLE, \textcolor{blue}{under (A2)} as $\cI\to\infty$, $\bExp[|S_{\tilde{\cT}}|]=o(\cI)$ if $e_k=o(\cI^2)$ such that $e_k\not=o(1)$, $\forall k$. And for U-sDMLE and U-dsDMLE, $\bExp[|S_{\tilde{\cT}}|]=o(\cI)$ if $T_U=o(\cI^2)$.
\end{lem}

\begin{IEEEproof}
The proof is presented in Appendix G.
\end{IEEEproof}

For the same reason stated below Lemma \ref{lem:appA} we can paraphrase the condition for LT-sDMLE and LT-dsDMLE in Lemma \ref{lem:appC} as $e_k=o(\cI^2)$ for practical purposes.
Now, we proceed to analyze the singly sequential estimator, LT-sDMLE.

\begin{thm}
    \label{thm:fade1}
    Consider the sequential decentralized estimator LT-sDMLE, given in Section \ref{sec:lt-sdmle}. \textcolor{blue}{Assuming (A1) and (A2)} it is, as $\cI\to\infty$, asymptotically unbiased, i.e., $\bExp[\tilde{x}_{\tilde{\cT}}-x]\to0$, and consistent, i.e., $\tilde{x}_{\tilde{\cT}} \overset{p}{\rightarrow} x$, if $R_k\to\infty$ at any rate, and $e_k\to\infty$ at a slower rate than $\cI$, i.e., $e_k=o(\cI)$, $\forall k$.
\end{thm}

\begin{IEEEproof}
    The proof is provided in Appendix H.
\end{IEEEproof}

\begin{thm}
    \label{thm:fade2}
    The sequential decentralized estimator LT-sDMLE, given in Section \ref{sec:lt-sdmle}, is, \textcolor{blue}{under (A1) and (A2)}, as $\cI\to\infty$, asymptotically optimal, i.e., $\sqrt{I_{\tilde{\cT}}^c} (\tilde{x}_{\tilde{\cT}}-x) \overset{d}{\rightarrow} \cN(0,1)$, if $R_k\to\infty$ at a faster rate than $\log \cI$, i.e., $R_k=\omega(\log \cI)$, $e_k\to\infty$ such that $e_k=o(\sqrt{\cI})$, and $r_U=\omega(\log(\sqrt{\cI}/e_k)),~\forall k$.
\end{thm}

\begin{IEEEproof}
    The proof is given in Appendix I.
\end{IEEEproof}

Next, we analyze LT-dsDMLE, in which, in addition to $\tilde{U}_{\tilde{\cT}}$, $\tilde{V}_{\tilde{\cT}}$ is also sequentially transmitted, as opposed to LT-sDMLE.

\begin{thm}
    \label{thm:fade3}
    Consider the sequential decentralized estimator LT-dsDMLE, given in Section \ref{sec:lt-dsdmle}. \textcolor{blue}{Under (A1) and (A2)} it is, as $\cI\to\infty$, asymptotically unbiased, i.e., $\bExp[\tilde{x}_{\tilde{\cT}}-x]\to0$, and consistent, i.e., $\tilde{x}_{\tilde{\cT}} \overset{p}{\rightarrow} x$, if $d_k\to\infty$, and $e_k\to\infty$ at slower rates than $\cI$, i.e., $d_k=o(\cI)$ and $e_k=o(\cI)$, $\forall k$.
\end{thm}

\begin{IEEEproof}
    The proof can be found in Appendix J.
\end{IEEEproof}

\begin{thm}
    \label{thm:fade4}
    \textcolor{blue}{With (A1) and (A2)} the sequential decentralized estimator LT-dsDMLE, given in Section \ref{sec:lt-dsdmle}, is, as $\cI\to\infty$, asymptotically optimal, i.e., $\sqrt{I_{\tilde{\cT}}^c} (\tilde{x}_{\tilde{\cT}}-x) \overset{d}{\rightarrow} \cN(0,1)$, if $d_k=o(\sqrt{\cI})$, $r_V=\omega(\log(\sqrt{\cI}/d_k))$, $e_k\to\infty$ such that $e_k=o(\sqrt{\cI})$, and $r_U=\omega(\log(\sqrt{\cI}/e_k)),~\forall k$.
\end{thm}

\begin{IEEEproof}
The proof is presented in Appendix K.
\end{IEEEproof}

Finally, in the following two theorems, we analyze U-sDMLE and U-dsDMLE, that are based on the conventional uniform sampling.

\begin{thm}
    \label{thm:uni1}
    \textcolor{blue}{Assuming (A1) and (A2)} the sequential decentralized estimator U-sDMLE, given in Section \ref{sec:u-sdmle} is, as $\cI\to\infty$, asymptotically unbiased, i.e., $\bExp[\tilde{x}_{\tilde{\cT}}-x]\to0$, and consistent, i.e., $\tilde{x}_{\tilde{\cT}} \overset{p}{\rightarrow} x$, if $r_U\to\infty$ at any rate, $T_U=O(\cI)$, and $R_k\to\infty$ at any rate. Moreover, it is asymptotically optimal, i.e., $\sqrt{I_{\tilde{\cT}}^c} (\tilde{x}_{\tilde{\cT}}-x) \overset{d}{\rightarrow} \cN(0,1)$, if $r_U=\omega(\log \cI)$, $T_U=o(\cI)$, and $R_k=\omega(\log \cI)$.
\end{thm}

\begin{IEEEproof}
The proof is provided in Appendix L.
\end{IEEEproof}

\begin{thm}
    \label{thm:uni2}
    \textcolor{blue}{With (A1) and (A2)} the sequential decentralized estimator U-dsDMLE, given in Section \ref{sec:u-dsdmle} is, as $\cI\to\infty$, asymptotically unbiased, i.e., $\bExp[\tilde{x}_{\tilde{\cT}}-x]\to0$, and consistent, i.e., $\tilde{x}_{\tilde{\cT}} \overset{p}{\rightarrow} x$, if $r_U\to\infty$ and $r_V\to\infty$ at any rate, and $T_U=O(\cI)$, $T_V=o(\cI)$. Moreover, it is asymptotically optimal, i.e., $\sqrt{I_{\tilde{\cT}}^c} (\tilde{x}_{\tilde{\cT}}-x) \overset{d}{\rightarrow} \cN(0,1)$, if $r_U=\omega(\log \cI)$, $r_V=\omega(\log \cI)$, $T_U=O(\cI)$, and $T_V=o(\sqrt{\cI})$.
\end{thm}

\begin{IEEEproof}
The proof is given in Appendix M.
\end{IEEEproof}

In order to make fair comparisons between the level-triggered-sampling-based estimators and the uniform-sampling-based estimators, we make the average message rates equal. Specifically, in the uniform-sampling-based estimators the average message rates are $\frac{K}{T_U}$ and $\frac{K}{T_V}$ since the FC receives $K$ messages every $T_U$ and $T_V$ units of time, respectively. For the level-triggered-sampling-based estimators, we are interested in computing the limits $\lim_{t\to\infty} \frac{N_t}{t}$ and $\lim_{t\to\infty} \frac{M_t}{t}$ as the average message rates, where $N_t$ and $M_t$ denote the numbers of messages received by the FC until time $t$ for $V_t$ and $U_t$, respectively. From \cite[Eq. (40)]{Yilmaz11}, we can write $\lim_{t\to\infty} \frac{N_t}{t}=\sum_{k=1}^K \frac{1}{\Exp[t_{1,V}^k]}$ and $\lim_{t\to\infty} \frac{M_t}{t}=\sum_{k=1}^K \frac{1}{\Exp[t_{1,U}^k]}$. Hence, we select the thresholds $\{d_k\}$ and $\{e_k\}$ so that $\sum_{k=1}^K \frac{1}{\Exp[t_{1,V}^k]}=\frac{K}{T_V}$ and $\sum_{k=1}^K \frac{1}{\Exp[t_{1,U}^k]}=\frac{K}{T_U}$, respectively.

The average number of bits transmitted by sensor $k$ until the stopping time represents the energy consumption and is given by $R_k=\Exp[N_{\tilde{\cT}}^k]r_V$ and $R_{k,U}=\Exp[M_{\tilde{\cT}}^k]r_U$. It was shown in Section \ref{sec:perf_awgn} that $R_k=\Theta\Big(\frac{\cI r_V}{d_k}\Big)$ as $\cI\to\infty$. Here in a similar fashion, using Lemma \ref{lem:appA}, we can verify the former for LT-dsDMLE and also show that $R_{k,U}=\Theta\Big(\frac{\cI r_U}{e_k}\Big)$ as $\cI\to\infty$ for LT-sDMLE and LT-dsDMLE. On the other hand, $N_{\tilde{\cT}}^k=\left\lfloor \frac{\tilde{\cT}}{T_V} \right\rfloor$ for U-dsDMLE, and $M_{\tilde{\cT}}^k=\frac{\tilde{\cT}}{T_U}$ for U-sDMLE and U-dsDMLE. Using again Lemma \ref{lem:appA}, as $\cI\to\infty$, we write $R_k=\Theta\Big(\frac{\cI r_V}{T_V}\Big)$ for U-dsDMLE, and $R_{k,U}=\Theta\Big(\frac{\cI r_U}{T_U}\Big)$ for U-sDMLE and U-dsDMLE.

\begin{table}[t]
  \centering
  \begin{tabular}{|p{2cm}|p{3.5cm}|p{3.5cm}|}
  \hline
  & \centering{Energy} & \hspace{1cm}Bandwidth \\
  \hline
  LT-sDMLE \newline (Thm. \ref{thm:fade1}~\&~\ref{thm:fade2})& AU\&C:~~$E\cong O(1)$ \newline AO:~~$E\cong \Theta(\sqrt{\cI})$ & AU\&C:~~$\bar{B}\cong O(1)$ \newline AO:~~$\bar{B}\cong \Theta(\log \cI)$ \\
  \hline
  LT-dsDMLE \newline (Thm. \ref{thm:fade3}~\&~\ref{thm:fade4})& AU\&C:~~$E\cong O(1)$ \newline AO:~~$E\cong \Theta(\sqrt{\cI})$ & AU\&C:~~$B=O(1)$ \newline AO:~~$B\cong O(1)$ \\
  \hline
  U-sDMLE \newline (Thm. \ref{thm:uni1})& AU\&C:~~$E\cong O(1)$ \newline AO:~~$E\cong \Theta(\log \cI)$ & AU\&C:~~$\bar{B}\cong O(1)$ \newline AO:~~$\bar{B}\cong \Theta(\log \cI)$ \\
  \hline
  U-dsDMLE \newline (Thm. \ref{thm:uni2})& AU\&C:~~$E\cong O(1)$ \newline AO:~~$E\cong \Theta(\sqrt{\cI}\log \cI)$ & AU\&C:~~$B\cong O(1)$ \newline AO:~~$B\cong \Theta(\log \cI)$ \\
  \hline
  \multicolumn{3}{|c|}{AU\&C:~asymptotic unbiasedness and consistency,~~AO:~asymptotic optimality} \\
  \hline
  \end{tabular}
  \caption{Practical implications of Theorems \ref{thm:fade1}--\ref{thm:uni2}.}
  \label{tab:fade}
\end{table}

\textcolor{blue}{In Table \ref{tab:fade}, using Theorems \ref{thm:fade1}--\ref{thm:uni2} we show the slowest (almost) achievable growth rates for the total energy and bandwidth consumptions to ensure asymptotic unbiasedness/consistency and asymptotic optimality. The total energy and bandwidth consumed at sensor $k$ is the summation of those consumed for the local processes $U_t^k$ and $V_t^k$. It is seen in the first two rows of Table \ref{tab:fade} that LT-sDMLE, which is a mixture of the sequential and fixed-time approaches, never consumes lees energy and bandwidth than LT-dsDMLE, which solely follows the sequential approach. Although the singly sequential U-sDMLE asymptotically consumes less energy than the doubly sequential U-dsDMLE, it is not a practical choice due to its non-asymptotical high bandwidth usage (cf. the last two rows of Table \ref{tab:fade}). Considering the (non-asymptotical) low bandwidth usage and the other practical advantages (cf. Section \ref{sec:awgn_alg}) of the sequential approach over the fixed-time approach, LT-dsDMLE and U-dsDMLE are in practice preferable to LT-sDMLE and U-sDMLE, respectively.}

We will focus on comparing the doubly sequential schemes LT-dsDMLE and U-dsDMLE due to the reasons stated above. LT-dsDMLE achieves asymptotic unbiasedness and consistency by keeping its bandwidth usage constant and increasing its energy usage at an arbitrarily slow rate, whereas U-dsDMLE needs to increase both its bandwidth and energy usages at arbitrarily slow rates. The superiority of the level-triggered-sampling-based LT-dsDMLE over the uniform-sampling-based U-dsDMLE is more obvious in achieving asymptotic optimality. As seen in the second and fourth rows of Table \ref{tab:fade}, for asymptotic optimality LT-dsDMLE needs to consume significantly less energy and bandwidth than U-dsDMLE needs, i.e., $\Theta(\sqrt{\cI})$ vs. $\Theta(\sqrt{\cI}\log\cI)$ and $O(1)$ vs. $\Theta(\log\cI)$.

This is because increasing the average sampling intervals $T_U$ and $T_V$, i.e., increasing the sampling thresholds $d_k$ and $e_k$ in LT-dsDMLE, without increasing $r_U$ and $r_V$, the numbers of transmitted bits  per sample, does help to improve the asymptotic performance of LT-dsDMLE. However, it does not help in U-dsDMLE.

The underlying reason for this fundamental difference is that the quantization errors in LT-dsDMLE, based on level-triggered sampling, are bounded constants that do not depend on the average sampling intervals. Hence, they become negligible compared to the messages $\tilde{v}_n^k$ and $\tilde{u}_n^k$, given by \eqref{eq:VincrRecov} and \eqref{eq:UincrRecov}, respectively, as the average sampling intervals tend to infinity, i.e., the thresholds $d_k$ and $e_k$ tend to infinity. On the other hand, in U-dsDMLE, based on uniform sampling, the quantization subintervals, and thus the quantization errors gets larger as the sampling periods $T_U$ and $T_V$ tend to infinity with constant $r_U$ and $r_V$.
As a result, the level-triggered-sampling-based LT-dsDMLE is in practice preferable to the uniform-sampling-based U-dsDMLE since it asymptotically requires considerably less energy and bandwidth than its competitor does.

\section{Simulation Results}
\label{sec:sim}

The asymptotic performances of the proposed decentralized estimators were analyzed in Section \ref{sec:perf}.
In this section, we provide simulation results to compare their non-asymptotic performances. Throughout the section, we use $r_V=r_U=1$ to illustrate the case of most practical interest, i.e., to conform to the low bandwidth usage requirement in decentralized systems. The thresholds $\{d_k\}$ and $\{e_k\}$ are determined to satisfy the given average sampling intervals $T_V$ and $T_U$, respectively. The quantization parameters $\phi_k$ and $\theta_k$ are set as the 99-th percentiles of $\left|\frac{2\Re((y^k_{\tau})^*h^k_{\tau})}{\sigma_k^2}\right|$ and $\frac{2|h^k_{\tau}|^2}{\sigma_k^2}$, respectively.

For the AWGN case, our performance metric is the mean squared error (MSE), i.e., $\Exp[(\tilde{x}_{t_{\cI}}-x)^2]$. And we plot it against four common parameters of both the centralized and the decentralized estimators, namely the stopping time $t_{\cI}$, known to be deterministic; the number of sensors $K$; the signal-to-noise ratio (SNR) of the channel $\text{SNR}_k=\frac{|h_k|^2}{\sigma_k^2}$; and the bounding constant of the parameter to be estimated, i.e., $\cX$ where $|x|<\cX$. Note that $\cX$ represents the uncertainty level in $x$, and affects the value of $\phi_k$, which defines the quantization intervals for $V_{t_{\cI}}^k$ in DMLE and $q_n^k$ in LT-DMLE.

On the other hand, for the fading case we use the expected stopping time $\Exp[\tilde{\cT}]$, as the performance metric. \textcolor{blue}{Rayleigh fading channel gains are used in the simulations, i.e., $\Re(h_t^k),\Im(h_t^k)\sim \cN(0,\sigma_h^2/2)$}.We plot $\Exp[\tilde{\cT}]$ against MSE, $K$, $\text{SNR}_k=\frac{\sigma_h^2}{\sigma_k^2}$, and $\cX$.

\subsection{AWGN Channels}

\textit{\underline{Fixed $K$, $\text{SNR}_k$, and $\cX$, varying $t_{\cI}$}}:
Firstly, we set $K=5$, $\text{SNR}_k=1$ ($0$ dB) $\forall k$, $\cX=5$, and vary $\cI=25\times2^m$ where $m=0,\ldots,5$. Then, from \eqref{eq:stoptime_awgn} we have $t_{\cI}=\Big\lceil \frac{\cI}{2~K~\text{SNR}} \Big\rceil=3,5,10,20,40,80$. We also increase the average sampling interval $T_V$ as the stopping time increases to meet the low communication rate requirement, i.e., $T_V=\Exp[t_{n,V}^k]=2\times1.4^m,~\forall k$. Recalling that $T_V=\Theta(d_k)$ (cf. the proof of Theorem \ref{thm:fade3}), we see that the rate of $T_V$ complies with Theorem \ref{thm:awgn4}, and also Theorem \ref{thm:awgn1} (cf. the discussion at the end of Section \ref{sec:perf_awgn}). In other words, the rate of $T_V$ (resp. $d_k,\forall k$), which is $1.4$, is smaller than but close to the rate of $\sqrt{\cI}$, which is $\sqrt{2}$. We keep the number of communication bits constant ($r_V=1$) in accordance with Theorem \ref{thm:awgn1} and Theorem \ref{thm:awgn4}. Hence, we maximize the performances of DMLE and LT-DMLE while conforming to the low communication rate and low bandwidth usage requirements.

\begin{figure}
\centering
\includegraphics[scale=0.7]{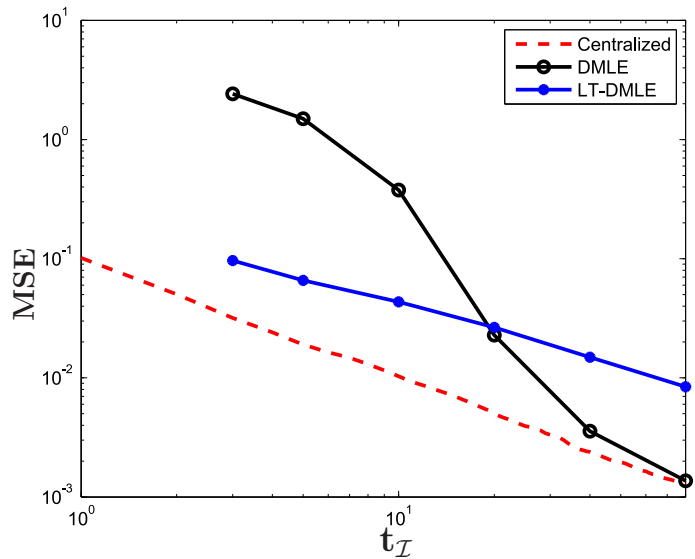}
\caption{Mean squared error (MSE), i.e., $\Exp[(\tilde{x}_{t_{\cI}}-x)^2]$, vs. the stopping (estimation) time, i.e., $t_{\cI}$, for the optimal centralized estimator, DMLE, and LT-DMLE with $r_V=1$.}
\label{fig:awgn_t}
\end{figure}

In Fig. \ref{fig:awgn_t}, it is seen that with short stopping times (up to $t_{\cI}=20$) LT-DMLE, following the sequential approach based on level-triggered sampling, performs significantly better than DMLE, that follows the fixed-time approach. However, when the stopping time becomes longer, DMLE outperforms LT-DMLE, and even reaches the optimal centralized estimation performance at $t_{\cI}=80$. This is due to the fact that the number of bits DMLE uses to quantize $V_{t_{\cI}}^k$, i.e., $R_k=\Exp[N_{t_{\cI}}^k]~r_V$, increases as the stopping time $t_{\cI}$ increases since the average number of messages transmitted in LT-DMLE until $t_{\cI}$, i.e., $\Exp[N_{t_{\cI}}^k]$, increases with increasing $t_{\cI}$. Accordingly, after $t_{\cI}=80$, $R_k$ becomes large enough that $V_{t_{\cI}}^k$ is fully recovered at the FC, i.e., $\tilde{V}_{t_{\cI}}^k=V_{t_{\cI}}^k$. In other words, the decentralized DMLE becomes the optimal centralized estimator. As pointed out in Section \ref{sec:awgn_alg}, DMLE does not meet the low bandwidth usage requirement, whereas LT-DMLE conforms to it by sending only $1$ bit in each sampling instant. Furthermore, DMLE provides no early estimates, whereas LT-DMLE does.

\textit{\underline{Fixed $t_{\cI}$, $\text{SNR}_k$, and $\cX$, varying $K$}}:
Secondly, we set $t_{\cI}=15$, $T_V=5$, $\text{SNR}_k=0~\text{dB},~\forall k$, $\cX=5$, and vary $K=2,\ldots,10$.
\begin{figure}
\centering
\includegraphics[scale=0.7]{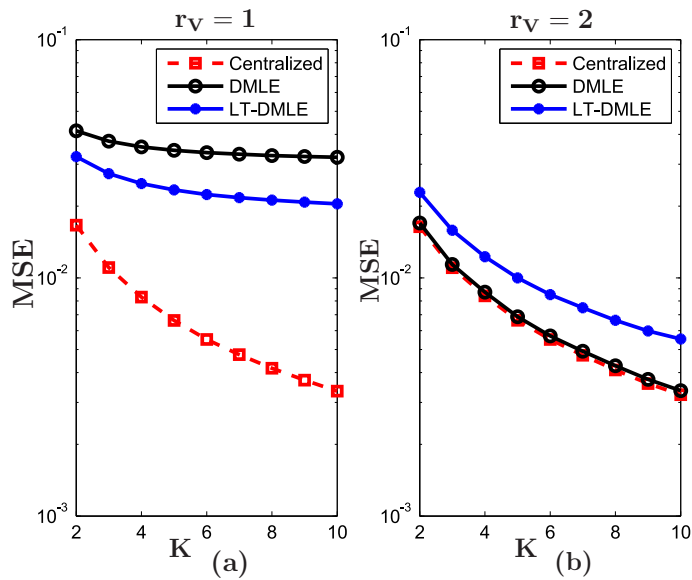}
\caption{Mean squared error (MSE), i.e., $\Exp[(\tilde{x}_{t_{\cI}}-x)^2]$, vs. the number of sensors, i.e., $K$, for the optimal centralized estimator, DMLE, and LT-DMLE with (a) $r_V=1$, (b) $r_V=2$.}
\label{fig:awgn_K}
\end{figure}
We plot the MSE vs. $K$ with $r_V=1$ and $r_V=2$ in Fig. \ref{fig:awgn_K}--a and Fig. \ref{fig:awgn_K}--b, respectively. With $r_V=1$, the case of most practical interest, it is seen that the optimal centralized estimator has an MSE decaying with rate $1/K$, but DMLE and LT-DMLE, the latter being superior, have MSEs decaying with rates slower than $1/K$. The quantization error (resp. overshoot) problem caused by small number of bits prevents DMLE (resp. LT-DMLE) from fully benefiting the increasing number of sensors.
However, when $r_V=2$, the MSE of both schemes seem to decay with rate $1/K$, as shown in Fig. \ref{fig:awgn_K}--b. In this case, DMLE, consuming high bandwidth at time $t_{\cI}$, attains the performance of the optimal centralized estimator.

\textit{\underline{Fixed $t_{\cI}$, $K$, and $\cX$, varying $\text{SNR}_k$}}:
Thirdly, we set $t_{\cI}=15$, $T_V=5$, $K=5$, $\cX=5$, and vary $\text{SNR}_k=-20,-10,\ldots,30~\text{dB},~\forall k$.
\begin{figure}
\centering
\includegraphics[scale=0.7]{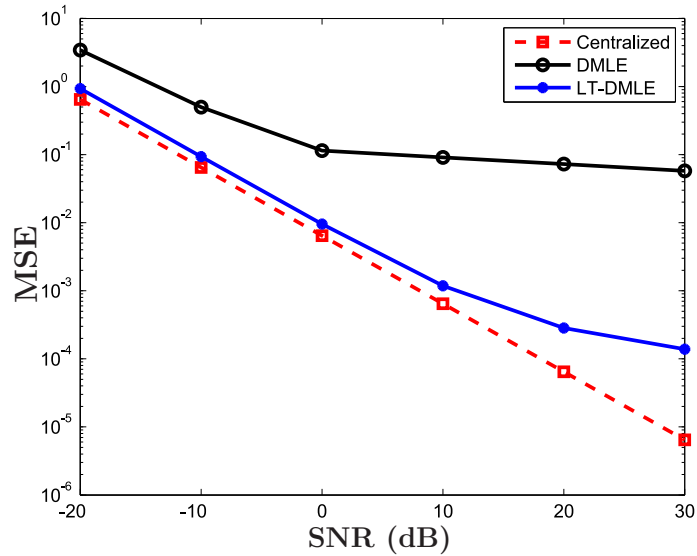}
\caption{Mean squared error (MSE), i.e., $\Exp[(\tilde{x}_{t_{\cI}}-x)^2]$, vs. SNR, i.e., $\frac{|h_k|^2}{\sigma_k^2}$, for the optimal centralized estimator, DMLE, and LT-DMLE with $r_V=1$.}
\label{fig:awgn_snr}
\end{figure}
In Fig. \ref{fig:awgn_snr}, it is seen that the MSEs of DMLE and LT-DMLE decay with decreasing rates as SNR increases, and even that of DMLE stops decreasing after SNR$=10$ dB. This is because the quantities to be transmitted to the FC, i.e., $V_{t_{\cI}}^k$ in DMLE and $v_n^k$ in LT-DMLE, take larger values as SNR increases. As a result, the quantization errors in DMLE with constant $R_k$ grows considerably, causing the improvement in the MSE performance to diminish. LT-DMLE is less affected by this phenomenon since via $1$ bit a significant part of $v_n^k$, i.e., $\tilde{v}_n^k=b_{n,1}^k d_k$ [cf. \eqref{eq:VincrRecov}], is transmitted in any case although the overshoot, $q_n^k$, grows with increasing SNR. Consequently, at high SNR LT-DMLE significantly outperforms DMLE.

\textit{\underline{Fixed $t_{\cI}$, $K$, and $\text{SNR}_k$, varying $\cX$}}:
Lastly, we set $t_{\cI}=15$, $T_V=5$, $K=5$, $\text{SNR}_k=0~\text{dB},~\forall k$, and vary $\cX=5\sqrt{10^m}$ where $m=-2,\ldots,2$.
\begin{figure}
\centering
\includegraphics[scale=0.7]{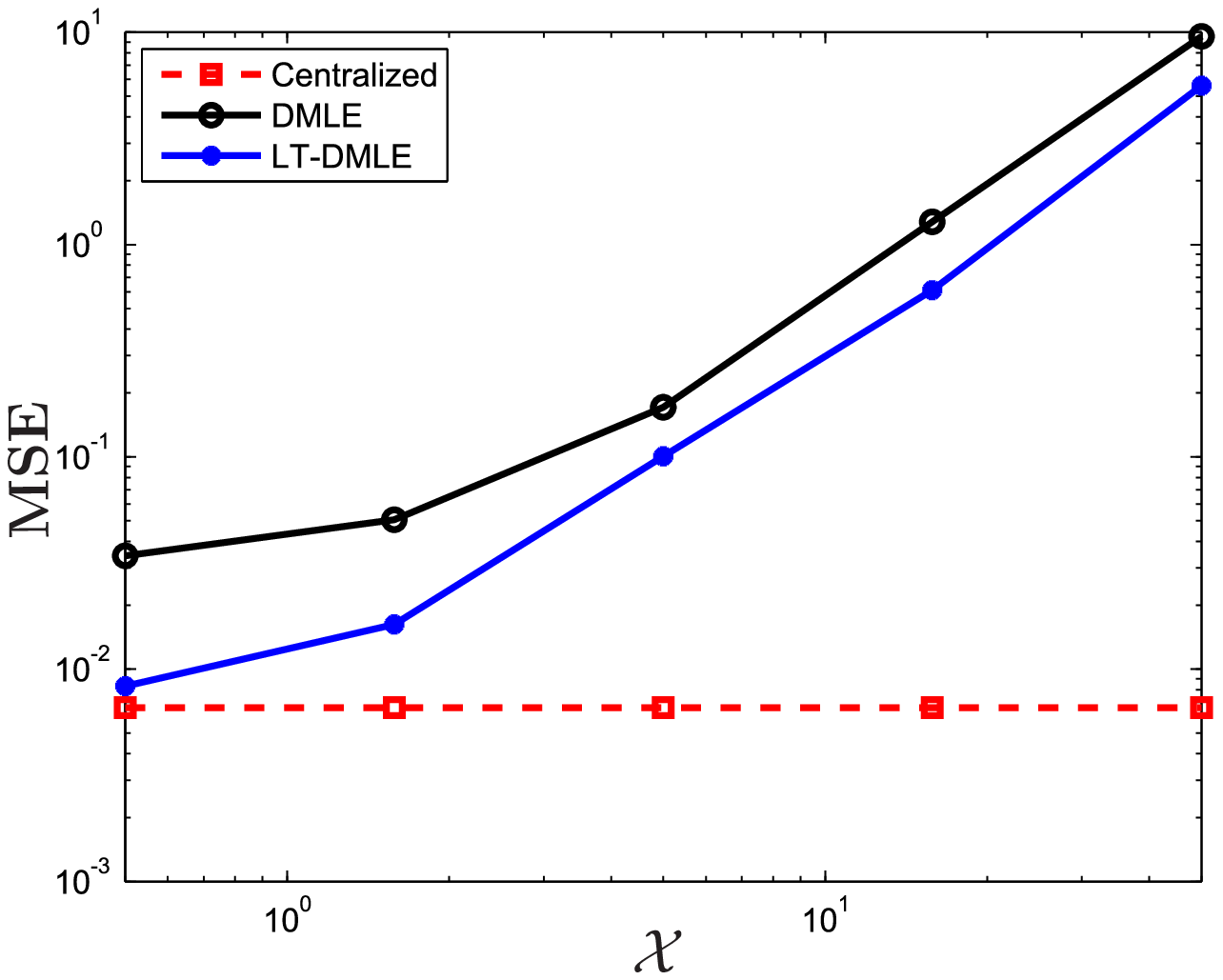}
\caption{Mean squared error (MSE), i.e., $\Exp[(\tilde{x}_{t_{\cI}}-x)^2]$, vs. the bounding constant of $x$, i.e., $\cX$, for the optimal centralized estimator, DMLE, and LT-DMLE with $r_V=1$.}
\label{fig:awgn_sx}
\end{figure}
It is seen in Fig. \ref{fig:awgn_sx} that the performance of the optimal centralized estimator is not affected by the increase in the uncertainty in $x$ since no quantization takes place, i.e., all local observations are available to the FC. On the other hand, those of DMLE and LT-DMLE, using constant number of bits, are deeply affected since quantization errors and overshoots grow with the increasing $\cX$, respectively. In LT-DMLE, with small values of $\cX$, e.g., $\cX=0.5$, the overshoot, $q_n^k$, is negligible compared to the magnitude of the transmitted value, $|\tilde{v}_n^k|=d_k$, hence we observe a performance close to the optimal one, and much better than that of DMLE. However, as $\cX$ increases, after $\cX=5$, $q_n^k$ dominates $|v_n^k|=d_k+q_n^k$, i.e., $q_n^k \gg d_k$, and thus the performance of LT-DMLE diverges from the optimal performance, and stays close to that of DMLE, which also diverges.

\subsection{Fading Channels}

Recall that under fading channels the sensor $k$ needs to transmit two random processes, namely, $U_t^k$ and $V_t^k$. The former should be sequentially transmitted since it determines the stopping time. Hence, we have two options to transmit $U_t^k$, namely, the conventional uniform sampler followed by a quantizer and the level-triggered sampler. On the other hand, we have three options for $V_t^k$ as it can also be transmitted non-sequentially (at once) at the stopping time. 

\textit{\underline{Fixed $K$, $\text{SNR}_k$, and $\cX$, varying MSE}}:
As in the AWGN case, we set $K=5$, $\text{SNR}_k=0 \text{dB},~\forall k$, $\cX=5$, and vary $\cI=25\times2^m$, $T_V=\Exp[t_{n,V}^k]=2\times1.4^m,~\forall k$ where $m=0,\ldots,5$. We also set $T_U=T_V$.

\begin{figure}
\centering
\includegraphics[scale=0.7]{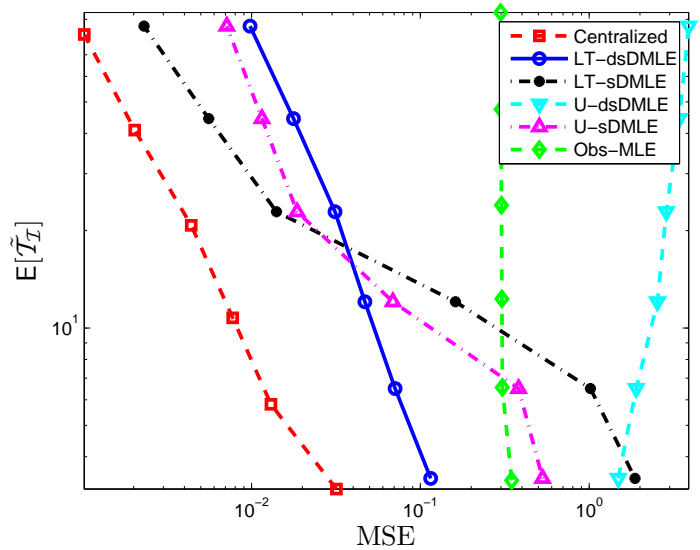}
\caption{Average stopping time, i.e., $\Exp[\tilde{\cT}]$, vs. MSE, i.e., $\Exp[(\tilde{x}_{t_{\cI}}-x)^2]$, for the optimal centralized estimator, the four decentralized estimators with $r_U=r_V=1$, and the sequential version of the scheme in \cite{Ribeiro06}.}
\label{fig:fade_mse}
\end{figure}

\textcolor{blue}{In this subsection, we compare the proposed decentralized estimators also with the sequential version of the estimator in \cite{Ribeiro06} \footnote{In \cite{Msechu12}, censoring is used with this estimator to decrease the energy and bandwidth consumption. Hence, the simulated estimator also corresponds to the uncensored version of the estimator in \cite{Msechu12}, whose MSE performance is obviously lower than its uncensored counterpart.}, in which the FC computes MLE using the one-bit quantized representations of sensor observations. Here we simulate the sequential version of that estimator where each sensor $k$, at each time $t$, transmits the one-bit representations of $\Re(y_t^k)$, $\Im(y_t^k)$, $\Re(h_t^k)$, and $\Im(h_t^k)$ [cf. \eqref{eq:observe}] until the stopping time given in \eqref{eq:Ustop}. Then, the FC computes MLE as
\be
\label{eq:obsMLE}
    \hat{x}_t = \frac{2\sigma}{\theta} \Phi^{-1}\left(\frac{N_t}{2Kt}\right),
\ee
where $\sigma$ is the variance of Gaussian noise at all sensors; $\pm\theta/2$ are the common quantization levels for $\{h_t^k\}$; $\Phi(\cdot)$ is the standard Gaussian cdf; $K$ is the number of sensors; and $N_t$ is the number of times the FC receives a pair $(\hat{y},\hat{h})$ such that $\frac{\hat{y}}{\hat{h}}>0$ until time $t$.
Note that this setup corresponds to $r_U=r_V=2$ and $T_U=T_V=1$, meaning that in this scheme sensors transmit more frequently more bits. Hence, the comparison between the proposed decentralized estimators and the observation-based MLE (Obs-MLE) in \eqref{eq:obsMLE} is in fact highly unfair in favor of the latter. Nevertheless, as shown in Fig. \ref{fig:fade_mse} the proposed doubly sequential level-triggered-sampling-based estimator (LT-dsDMLE) considerably outperforms Obs-MLE. This is because the MSE of Obs-MLE decreases very slowly as the average stopping time increases, e.g., $\text{MSE}=0.3470, 0.3059, 0.3039, 0.3019, 0.3007, 0.2990$ in Fig. \ref{fig:fade_mse}.}

In Fig. \ref{fig:fade_mse}, an important observation is the poor performance of U-dsDMLE, which uses uniform sampling to transmit $V_t^k$. This is because the local incremental process $v_{mT_V}^k$, which forms the $m$-th message from the sensor $k$, can take both negative and positive values, and with $r_V=1$ it cannot be accurately quantized. On the other hand, in LT-dsDMLE $r_V=1$ suffices to represent the local process $v_n^k$ well enough at the random sampling time $t_{n,V}^k$ [cf. \eqref{eq:samp}]. As a result, the doubly sequential LT-dsDMLE based on level-triggered sampling significantly outperforms the doubly sequential U-dsDMLE based on uniform sampling, which are of special interest to us as only the doubly sequential schemes enable low bandwidth usage.

The singly sequential schemes LT-sDMLE and U-sDMLE, using much higher bandwidth than their doubly sequential counterparts LT-dsDMLE and U-dsDMLE, improve their performance and outperform LT-dsDMLE after some point as the target MSE gets smaller. This is expected since LT-sDMLE and U-sDMLE use more and more bits (i.e., consume higher and higher bandwidth) to transmit $V_{\tilde{\cT}}^k$ as the stopping time $\tilde{\cT}$ grows. Hence, in fact, the comparison between the singly sequential schemes and the doubly sequential schemes is not completely fair. Note that here in the fading case, the performances of LT-sDMLE and U-sDMLE do not converge to that of the optimal centralized scheme (unlike DMLE in Fig. \ref{fig:awgn_t}) since $U_t^k$ is sequentially transmitted with a constant number of bits, $r_U=1$.

At moderate and high MSE values, we observe a compatibility problem in LT-sDMLE since a conventional quantizer is used to transmit $V_{\tilde{\cT}}^k$, whereas $U_{\tilde{\cT}}^k$ is transmitted via level-triggered sampling.
We observe such a problem since the decentralized estimates in this paper are computed as the ratio of $\tilde{V}_t$ to $\tilde{U}_t$, and when $\tilde{U}_t$ and $\tilde{V}_t$ are computed via different methods (i.e., one via a conventional quantizer and the other via level-triggered sampling), quantization errors in $\tilde{U}_t$ and $\tilde{V}_t$ are of different orders of magnitude.
Therefore, U-sDMLE, using conventional quantizers in transmitting both  $V_{\tilde{\cT}}^k$ and $U_{\tilde{\cT}}^k$ (although the latter is sequentially transmitted), performs better than LT-sDMLE at moderate and high MSE values. However, at low MSE values the singly sequential schemes practically transmit $V_{\tilde{\cT}}^k$ exactly (as the number of bits $R_k$ gets larger), eliminating the compatibility problem in LT-sDMLE, and thus LT-sDMLE outperforms U-sDMLE, demonstrating the superiority of level-triggered sampling over uniform sampling in another way.

\textit{\underline{Fixed MSE, $\text{SNR}_k$, and $\cX$, varying $K$}}:
\begin{figure}
\centering
\includegraphics[scale=0.7]{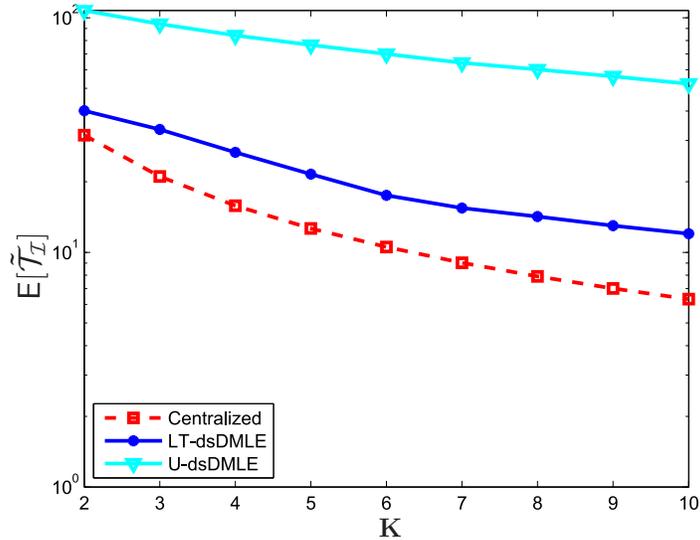}
\caption{Average stopping time, i.e., $\Exp[\tilde{\cT}]$, vs. the number of sensors, i.e., $K$, for the optimal centralized estimator, LT-dsDMLE, and U-dsDMLE with $r_U=1$, $r_V=2$.}
\label{fig:fade_K}
\end{figure}
Henceforth, for the sake of clarity and brevity, we will only consider the doubly sequential estimators, LT-dsDMLE and U-dsDMLE, since the singly sequential estimators, LT-sDMLE and U-sDMLE, violate the low bandwidth usage requirement. Next we set $\text{MSE}=10^{-2}$, $\text{SNR}_k=0 \text{dB},~\forall k$, $\cX=5$, and vary $K=2,\ldots,10$. To make the MSEs of the optimal centralized estimator, LT-dsDMLE and U-dsDMLE equal to the target value, the target Fisher information and the average sampling intervals, for each scheme, are determined as $\cI=25\times2^s$, and $T_U=T_V=\Exp[t_{n,V}^k]=2\times1.4^s,~\forall k$ where $s\in\bR$. Note that for each scheme $s$ takes different values in general. We will use $r_U=1$, as before, but $r_V=2$ from now on to enable U-dsDMLE to achieve the target MSE (see Fig. \ref{fig:fade_mse}).

As shown in Fig. \ref{fig:fade_K}, the average stopping time of the centralized scheme decays with a rate close to $1/K$, whereas those of LT-dsDMLE and U-dsDMLE, the former being faster, are slower than $1/K$ for the same reason as in the AWGN case. Recall that in the AWGN case $r_V=2$ was sufficient for the decentralized schemes to enjoy the increasing sensor diversity completely (see Fig. \ref{fig:awgn_K}-b). However, here under fading channels $r_V=2$, together with $r_U=1$, does not suffice to alleviate the quantization error problem to fully exploit the increasing sensor diversity.

\textit{\underline{Fixed MSE, $K$, and $\cX$, varying $\text{SNR}_k$}}:
We set $K=5$, and vary $\text{SNR}_k=-20,-10,\ldots,20~\text{dB},~\forall k$.
\begin{figure}
\centering
\includegraphics[scale=0.7]{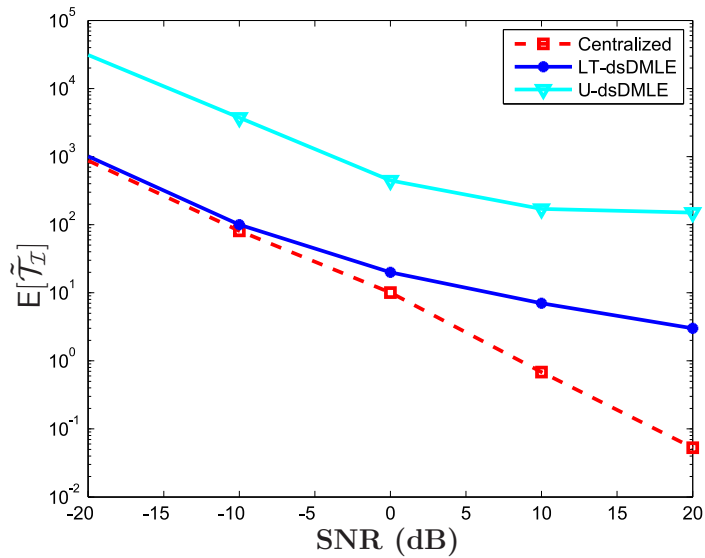}
\caption{Average stopping time, i.e., $\Exp[\tilde{\cT}]$, vs. SNR, i.e., $\frac{\Exp[|h_t^k|^2]}{\sigma_k^2}$, for the optimal centralized estimator, LT-dsDMLE, and U-dsDMLE with $r_U=1$, $r_V=2$.}
\label{fig:fade_snr}
\end{figure}
It is seen in Fig. \ref{fig:fade_snr} that the average stopping times of the centralized estimator, LT-dsDMLE and U-dsDMLE decrease with the increasing SNR, as expected, but the rates of
LT-dsDMLE and U-dsDMLE slow down for the same reason as in the AWGN case. The quantities to be transmitted become larger as SNR increases, hence with constant $r_U$ and $r_V$ the quantization errors and overshoots get larger, slowing down the performance improvement. We observe that the average stopping time of U-dsDMLE is likely to stop decreasing after $20$ dB, whereas that of LT-dsDMLE continues to decrease since the rate of increase of the overshoots in this case is slower than that of the quantization errors in U-dsDMLE, demonstrating another advantage of LT-dsDMLE over U-dsDMLE. Specifically, U-dsDMLE quantizes $u_{mT_U}^k\in[0,T_U\theta_k]$ and $v_{mT_V}^k\in[-T_V\phi_k,T_V\phi_k]$ where $\phi_k$ increases with the increasing SNR. On the other hand, the overshoots in LT-dsDMLE are confined to $[0,\theta_k]$ or $[0,\phi_k]$ with a high probability (cf. Section \ref{sec:alg}).

\textit{\underline{Fixed MSE, $K$, and $\text{SNR}_k$, varying $\cX$}}:
Lastly, we vary $\cX=5\sqrt{10^m}$ where $m=-2,\ldots,2$, setting the other parameters to the same values used in the previous subsections.
\begin{figure}
\centering
\includegraphics[scale=0.7]{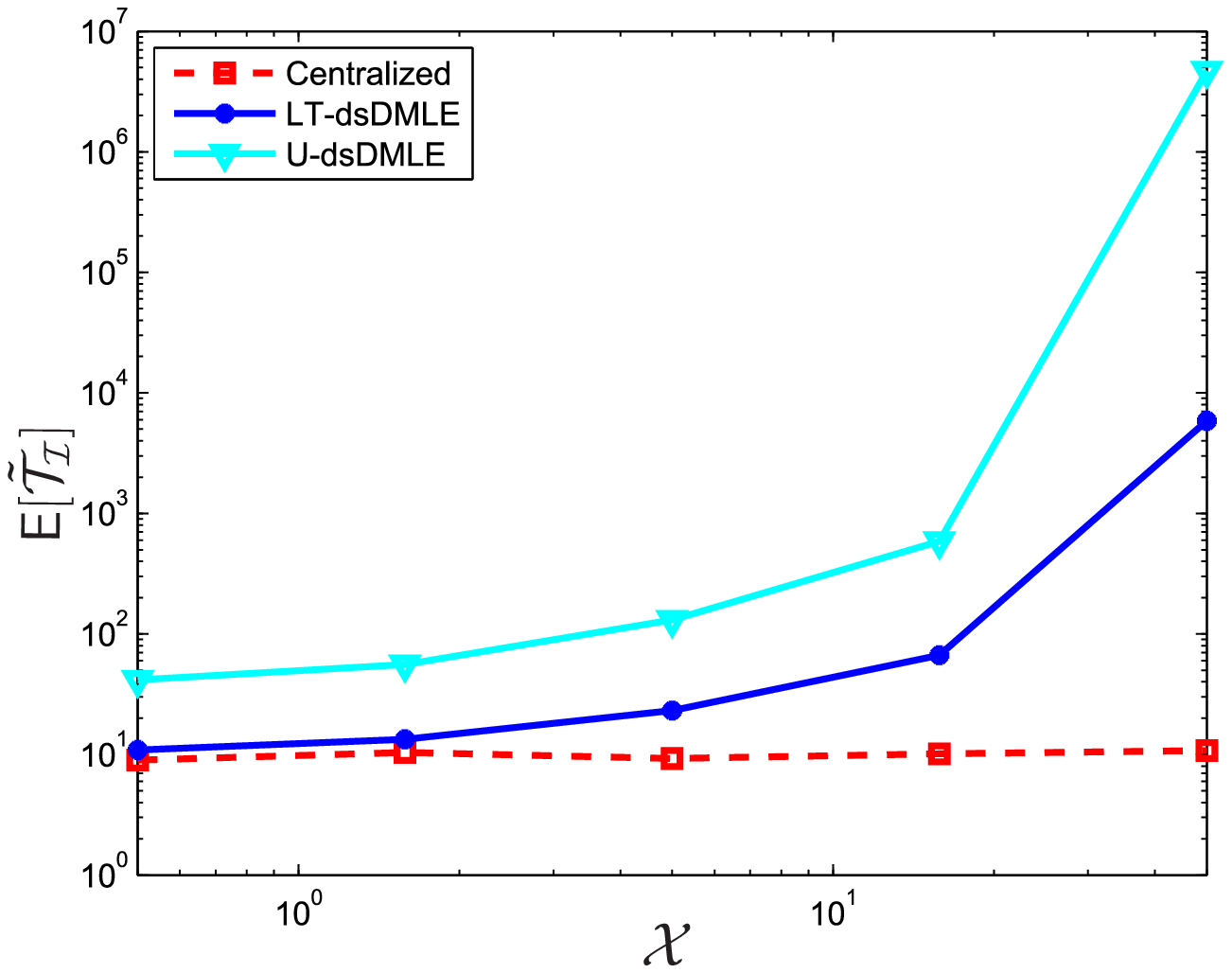}
\caption{Average stopping time, i.e., $\Exp[\tilde{\cT}]$, vs. the bounding constant of $x$, i.e., $\cX$, for the optimal centralized estimator, LT-dsDMLE, and U-dsDMLE with $r_U=1$, $r_V=2$.}
\label{fig:fade_sx}
\end{figure}
Fig. \ref{fig:fade_sx} shows that the average stopping times of the decentralized schemes diverge from that of the centralized scheme as $\cX$ increases since the overshoots and the quantization errors grow with increasing $\cX$ in LT-dsDMLE and U-dsDMLE, respectively, as described in the AWGN case. In particular, we observe that increasing $\cX$ causes $\phi_k$ to grow, hence as explained in the previous subsection the quantization errors in U-dsDMLE grow much faster than the overshoots in LT-dsDMLE as $\cX$ increases. Accordingly, U-dsDMLE diverges much quicker than LT-dsDMLE, as shown in Fig. \ref{fig:fade_sx}.

\section{Conclusion}
\label{sec:conc}

We have proposed and rigorously analyzed a new decentralized estimation framework based on a non-uniform sampling technique, namely level-triggered sampling. Level-triggered sampling, eliminating the need for quantization, produces a single bit, and thus provides an efficient way of information transmission in decentralized systems. It is used in the proposed estimator to effectively report local observations at sensors to a fusion center (FC). Messages received from sensors are combined at the FC to compute approximation(s) to global random process(es) that characterize(s) the centralized maximum likelihood estimator (MLE), shown to be optimal. Performing an asymptotic analysis we have determined sufficient conditions under which the proposed estimator and the decentralized estimator based on conventional uniform sampling are asymptotically unbiased, consistent and asymptotically optimal. In particular, it is sufficient for the proposed estimator to have average communication (sampling) intervals tending to infinity at rates lower than specific upper bounds, and transmit a constant number of bits at each communication time. On the other hand, for the scheme based on uniform sampling the number of bits transmitted at each communication time has to tend to infinity at rates faster than specific lower bounds, regardless of the average communication intervals. For low bandwidth and energy usage it is practically desired to have the number of bits as small as possible, and the average communication intervals as large as possible. In that aspect, the analytical results clearly demonstrates the superiority of the proposed scheme over the conventional scheme. Simulation results further demonstrate the superior non-asymptotic performance of the proposed scheme based on level-triggered sampling under different conditions. \textcolor{blue}{In a future work we plan to consider the case of noisy transmission channels between sensors and the FC, as in \cite{Yilmaz13}, which studies the decentralized detection problem.}


\section*{Appendix A: Proof of Lemma \ref{lem:cen}}

From \eqref{eq:MLEdist} it is seen that $\hat{x}_t$ is conditionally unbiased. Consistency and efficiency follow from \eqref{eq:MLEdist} and \eqref{eq:Finf}. We have $\Exp[(\hat{x}_t-x)^2 | \cH_t]=\text{Var}(\hat{x}_t | \cH_t)=1/U_t=1/I_t^c$, i.e., efficiency. If we have $U_t\overset{a.s.}{\rightarrow}\infty$, i.e., $\Pro(\lim_{t\to\infty} \sum_{k=1}^K \sum_{\tau=1}^t \frac{2|h^k_{\tau}|^2}{\sigma_k^2}=\infty)=1$, then $\hat{x}_t \overset{L^2}{\rightarrow} x$ implying $\hat{x}_t \overset{p}{\rightarrow} x$ given $\cH_t$, i.e., consistency. If $\Pro(\lim_{t\to\infty} \sum_{k=1}^K \sum_{\tau=1}^t \frac{2|h^k_{\tau}|^2}{\sigma_k^2}=\infty)\not=1$, then there exists some $M<\infty$ such that $\Pro(\lim_{t\to\infty} \sum_{k=1}^K \sum_{\tau=1}^t \frac{2|h^k_{\tau}|^2}{\sigma_k^2}<M)\not=0$. Hence, it suffices to show that $\Pro(\lim_{t\to\infty} \sum_{k=1}^K \sum_{\tau=1}^t \frac{2|h^k_{\tau}|^2}{\sigma_k^2}<M)=0,~\forall M<\infty$. Note that
    \begin{align}
    \Pro\left(\lim_{t\to\infty} \sum_{k=1}^K \sum_{\tau=1}^t \frac{2|h^k_{\tau}|^2}{\sigma_k^2}<M \right) \leq& \lim_{t\to\infty} \Pro\left(\sum_{k=1}^K \sum_{\tau=1}^t \frac{2|h^k_{\tau}|^2}{\sigma_k^2}<M \right) \nn\\
    =& \lim_{t\to\infty} \Pro\left( \exp\left(-\sum_{k=1}^K \sum_{\tau=1}^t \frac{2|h^k_{\tau}|^2}{\sigma_k^2} \right)>\exp(-M) \right) \nn\\
    \leq& \lim_{t\to\infty} \frac{\left(\Exp\left[\exp\left(-\sum_{k=1}^K \frac{2|h^k_1|^2}{\sigma_k^2}\right) \right] \right)^t} {\exp(-M)}, \nn
    \end{align}
    where the last inequality follows from Markov's inequality and the fact that $\left\{\sum_{k=1}^K \frac{2|h^k_t|^2}{\sigma_k^2}\right\}_t$ are i.i.d.. Now, since $\sum_{k=1}^K \frac{2|h^k_1|^2}{\sigma_k^2}>0$ from (A2), $\exp(-\sum_{k=1}^K \frac{2|h^k_1|^2}{\sigma_k^2})<1$ and $\Exp[\exp(-\sum_{k=1}^K \frac{2|h^k_1|^2}{\sigma_k^2})]<1$. Hence, $\lim_{t\to\infty} (\Exp[\exp(-\sum_{k=1}^K \frac{2|h^k_1|^2}{\sigma_k^2})])^t=0$, concluding the proof.

\section*{Appendix B: Proof of Theorem \ref{thm:awgn1}}

    To prove the first part of the theorem it is sufficient to show that $\tilde{x}_{t_{\cI}} \overset{L^1}{\rightarrow} x$, i.e., $\Exp[|\tilde{x}_{t_{\cI}}-x|]\to0$, since $\tilde{x}_{t_{\cI}} \overset{L^1}{\rightarrow} x$ implies both $\tilde{x}_{t_{\cI}} \overset{p}{\rightarrow} x$, and $\Exp[\tilde{x}_{t_{\cI}}-x]\to0$.
    Since we have $\Exp[|\tilde{x}_{t_{\cI}}-x|]=\Exp[|\tilde{x}_{t_{\cI}}-\hat{x}_{t_{\cI}}+\hat{x}_{t_{\cI}}-x|] \leq \Exp[|\tilde{x}_{t_{\cI}}-\hat{x}_{t_{\cI}}|]+\Exp[|\hat{x}_{t_{\cI}}-x|]$, and $\Exp[|\hat{x}_{t_{\cI}}-x|]\to0$ implied by $\hat{x}_{t_{\cI}} \overset{L^2}{\rightarrow} x$ from Lemma \ref{lem:cen}, we only need to show that $\Exp[|\tilde{x}_{t_{\cI}}-\hat{x}_{t_{\cI}}|]\to0$.
    Using (\ref{eq:MLE}) and (\ref{eq:dmle2}) we write $|\tilde{x}_{t_{\cI}}-\hat{x}_{t_{\cI}}|=\frac{|\tilde{V}_{t_{\cI}}-V_{t_{\cI}}|}{U_{t_{\cI}}}$ as $U_{t_{\cI}}\geq0$ [cf. (\ref{eq:Udef})]. From \eqref{eq:Vdef} and \eqref{eq:dmle1} taking the expectations of both sides we have
    \be
        \label{eq:thm_aw1_1}
        \Exp\big[|\tilde{x}_{t_{\cI}}-\hat{x}_{t_{\cI}}|\big] \leq \frac{\sum_{k=1}^K \Exp\big[|\tilde{V}_{t_{\cI}}^k-V_{t_{\cI}}^k|\big]}{U_{t_{\cI}}}.
    \ee
    \textcolor{blue}{The quantizer in DMLE is designed so that $\Exp\big[|\tilde{V}_{t_{\cI}}^k-V_{t_{\cI}}^k|\big]<\frac{t_{\cI}\phi_k}{2^{R_k}}=\frac{\Theta(t_{\cI})}{2^{R_k}}$}. We also have $U_{t_{\cI}}=t_{\cI} \sum_{k=1}^K \frac{2|h_k|^2}{\sigma_k^2}=\Theta(t_{\cI})$ as it is assumed in (A2) that $0<|h_k|^2<\infty,~\forall k$. Hence, using \eqref{eq:thm_aw1_1} we write
    \be
        \label{eq:thm_aw1_2}
        \Exp\big[|\tilde{x}_{t_{\cI}}-\hat{x}_{t_{\cI}}|\big] < \sum_{k=1}^K \frac{O(1)}{2^{R_k}},
    \ee
where it is sufficient to have $R_k\to\infty,~\forall k$, at any rate for asymptotic unbiasedness and consistency.

    For asymptotic optimality, note that $I_{t_{\cI}}=U_{t_{\cI}}$, and we can write $\sqrt{U_{t_{\cI}}} (\tilde{x}_{t_{\cI}}-x)=\sqrt{U_{t_{\cI}}} (\tilde{x}_{t_{\cI}}-\hat{x}_{t_{\cI}}) + \sqrt{U_{t_{\cI}}} (\hat{x}_{t_{\cI}}-x)$. From \eqref{eq:MLEdist} we have $\sqrt{U_{t_{\cI}}} (\hat{x}_{t_{\cI}}-x) \sim \cN(0,1)$. Hence, it is sufficient to show that $\sqrt{U_{t_{\cI}}} (\tilde{x}_{t_{\cI}}-\hat{x}_{t_{\cI}})\to0$ as $\cI\to\infty$. From \eqref{eq:thm_aw1_2}, if $R_k=\omega(\log U_{t_{\cI}}),~\forall k$, then $\sqrt{U_{t_{\cI}}} \Exp\big[|\tilde{x}_{t_{\cI}}-\hat{x}_{t_{\cI}}|\big] \to 0$, and thus from Markov's inequality $\sqrt{U_{t_{\cI}}} |\tilde{x}_{t_{\cI}}-\hat{x}_{t_{\cI}}| \to 0$, which implies $\sqrt{U_{t_{\cI}}} (\tilde{x}_{t_{\cI}}-\hat{x}_{t_{\cI}})\to0$.
    We have from \eqref{eq:stoptime_awgn} that $\cI \leq U_{t_{\cI}} \leq \cI + \sum_{k=1}^K \frac{2|h_k|^2}{\sigma_k^2}$, implying that $U_{t_{\cI}}=\cI+O(1)$ due to (A2), and hence the result in Theorem \ref{thm:awgn1} follows.

\section*{Appendix C: Proof of Theorem \ref{thm:awgn3}}

    As stated in the proof of Theorem \ref{thm:awgn1}, it is sufficient to show that $\Exp[|\tilde{x}_{t_{\cI}}-\hat{x}_{t_{\cI}}|]\to0$.
    Note from \eqref{eq:Vdef} that $V_{t_{\cI}}^k = \sum_{n=1}^{N_{t_{\cI}}^k} v_n^k + \sum_{\tau=t_{N_{t_{\cI}}^k,V}^k+1}^{t_{\cI}} \frac{2\Re((y^k_{\tau})^*h^k_{\tau})}{\sigma_k^2}$, and from \eqref{eq:VRecov} that $\tilde{V}_{t_{\cI}}^k = \sum_{n=1}^{N_{t_{\cI}}^k} \tilde{v}_n^k$. Thus, following \eqref{eq:thm_aw1_1} we have
    \be
        \label{eq:thm_aw3_1}
        |\tilde{x}_{t_{\cI}}-\hat{x}_{t_{\cI}}|\leq \frac{\sum_{k=1}^K \sum_{n=1}^{N_{t_{\cI}}^k}|\tilde{v}_n^k-v_n^k|}{U_{t_{\cI}}} + \frac{\sum_{k=1}^K \Big|\sum_{\tau=t_{N_{t_{\cI}}^k,V}^k+1}^{t_{\cI}} \frac{2\Re((y^k_{\tau})^*h_k)}{\sigma_k^2}\Big|}{U_{t_{\cI}}},
    \ee
    where in the second term of the right hand-side we can write $\Big|\sum_{\tau=t_{N_{t_{\cI}}^k,V}^k+1}^{t_{\cI}} \frac{2\Re((y^k_{\tau})^*h_k)}{\sigma_k^2}\Big| < d_k$ since it is known that no sampling occurs between the last sampling time, $t_{N_{t_{\cI}}^k,V}^k$, and the stopping time, $t_{\cI}$. Taking the expectations of both sides in \eqref{eq:thm_aw3_1} and noting that $U_{t_{\cI}}={t_{\cI}}\sum_{k=1}^K\frac{2|h_k|^2}{\sigma_k^2}$ in the AWGN case, we write
    \be
        \label{eq:thm_aw3_2}
        \Exp[|\tilde{x}_{t_{\cI}}-\hat{x}_{t_{\cI}}|] < \frac{1}{\sum_{k=1}^K \frac{2|h_k|^2}{\sigma_k^2}} \Big( \sum_{k=1}^K \frac{\Exp[\sum_{n=1}^{N_{t_{\cI}}^k}|\tilde{v}_n^k-v_n^k|]}{t_{\cI}} + \sum_{k=1}^K \frac{d_k}{t_{\cI}} \Big),
    \ee
    where the term outside the parentheses is $O(1)$ as $\cI\to\infty$ due to (A2).
    In the first term inside the parentheses, $|\tilde{v}_n^k-v_n^k|$ is the quantization error in absolute value of the $n$-th message from sensor $k$. Noting that $v_n^k=\sum_{\tau=t_{n-1,V}^k+1}^{t_{n,V}^k} \frac{2\Re((y^k_{\tau})^*h_k)}{\sigma_k^2}$, we see that $|\tilde{v}_n^k-v_n^k|$ depends only on the observations in the $n$-th intersampling period, i.e., $\{y_{\tau}^k\},~\tau \in (t_{n-1,V}^k,t_{n,V}^k]$, and thus $\{|\tilde{v}_n^k-v_n^k|\}_n$ are i.i.d.. Hence, the term $\sum_{n=1}^{N_{t_{\cI}}^k}|\tilde{v}_n^k-v_n^k|$ in \eqref{eq:thm_aw3_2} is a renewal reward process.
    Note from \eqref{eq:stoptime_awgn} and (A2) that $t_{\cI}=\Theta(\cI)$, i.e., $t_{\cI}\to\infty$ as $\cI\to\infty$.
    Hence, from \cite[Theorem 3.6.1]{Ross96} we have
    \be
    \label{eq:ren_th}
        \frac{\Exp[\sum_{n=1}^{N_{t_{\cI}}^k}|\tilde{v}_n^k-v_n^k|]}{t_{\cI}} \to \frac{\Exp[|\tilde{v}_1^k-v_1^k|]}{\Exp[t_{1,V}^k]}
    \ee
    as $\cI\to\infty$, where $\Exp[t_{1,V}^k]$ is the average sampling interval of sensor $k$. Then, it is sufficient to show that
    \be
        \label{eq:thm_aw3_3}
        \frac{\Exp[|\tilde{v}_1^k-v_1^k|]}{\Exp[t_{1,V}^k]}\to0, \ \ \ \text{and} \ \ \ \frac{d_k}{t_{\cI}}\to0,~\forall k,
    \ee
    as $\cI\to\infty$. If $d_k\to\infty$ such that $d_k=o(\cI)$, i.e., $d_k=o(t_{\cI}),~\forall k$, then both conditions in \eqref{eq:thm_aw3_3} will be satisfied since \textcolor{blue}{$\Exp[|\tilde{v}_1^k-v_1^k|]=\Exp[|\tilde{q}_1^k-q_1^k|]<\frac{\phi_k}{2^{r_V}}=O(1)$ for $r_V=O(1)$ as a result of the quantizer design in LT-DMLE}, and $\Exp[t_{1,V}^k]\to\infty$ as $d_k\to\infty$ [cf. \eqref{eq:samp}] as shown in Appendix D, concluding the proof.

\section*{Appendix D: Proof of Theorem \ref{thm:awgn4}}

    As stated in the proof of Theorem \ref{thm:awgn1}, it is sufficient to show that $\sqrt{U_{t_{\cI}}} (\tilde{x}_{t_{\cI}}-\hat{x}_{t_{\cI}})\to0$ as $\cI\to\infty$.
    If we show that $\sqrt{U_{t_{\cI}}} \Exp[|\tilde{x}_{t_{\cI}}-\hat{x}_{t_{\cI}}|] \to 0$, then from Markov's inequality we will have $\sqrt{U_{t_{\cI}}} |\tilde{x}_{t_{\cI}}-\hat{x}_{t_{\cI}}|\to0$, which implies $\sqrt{U_{t_{\cI}}} (\tilde{x}_{t_{\cI}}-\hat{x}_{t_{\cI}})\to0$. From \eqref{eq:thm_aw3_2} and the discussion following it, we can write
    \be
        \label{eq:thm_aw4_1}
        \sqrt{U_{t_{\cI}}} \Exp[|\tilde{x}_{t_{\cI}}-\hat{x}_{t_{\cI}}|] < O(1) \Big( \sum_{k=1}^K \frac{\Exp[|\tilde{v}_1^k-v_1^k|] \sqrt{U_{t_{\cI}}}}{\Exp[t_{1,V}^k]} + \sum_{k=1}^K \frac{d_k}{\Theta(\sqrt{U_{t_{\cI}}})} \Big)
    \ee
    as $\cI\to\infty$. Since $\Big\{\frac{2\Re((y^k_{\tau})^*h_k)}{\sigma_k^2}\Big\}_{\tau}$ are i.i.d., using Wald's identity we can write $\Exp[v_1^k]=\Exp[t_{1,V}^k]\Exp\Big[\frac{2\Re((y^k_1)^*h_k)}{\sigma_k^2}\Big]$. Since $\Exp\Big[\frac{2\Re((y^k_1)^*h_k)}{\sigma_k^2}\Big]=\frac{2x}{\sigma_k^2}|h_k|^2$, from (A1) and (A2) it is $O(1)$. At each sampling time, $v_n^k$ either crosses $d_k$ or $-d_k$, hence its expectation is given by $\Exp[v_1^k]=(1-\alpha_k)(d_k+\Exp[q_1^k|v_1^k\geq d_k])+\alpha_k(-d_k-\Exp[q_1^k|v_1^k\leq -d_k])=(1-2\alpha_k)d_k+\Exp[q_1^k]$, where $\alpha_k\triangleq \Pro(v_n^k\leq-d_k)$, and  $q_1^k$ is the overshoot bounded by $\frac{2}{\sigma_k^2}\big|\Re((y^k_{t_1^k})^*h_k)\big|$ (cf. Section \ref{sec:lt-dmle}). \textcolor{blue}{We have
    \be
    \label{eq:bdd_obs}
    \Exp\big[|\Re((y^k_t)^*h_k)|\big]\leq |x| |h_k|^2 + \Exp\big[ |\Re(w_1^k)| \big] |\Re(h_k)| + \Exp\big[ |\Im(w_1^k)| \big] |\Im(h_k)|,
    \ee
    where $\Exp\big[ |\Re(w_1^k)| \big]=\Exp\big[ |\Im(w_1^k)| \big]=\frac{\sigma_k}{\sqrt{\pi}}$, hence $\Exp[q_1^k]=O(1)$ from (A1) and (A2). Therefore, we have $\Exp[v_1^k]=\Theta(d_k)$ and $\Exp[t_{1,V}^k]=\Theta(d_k)$. Note that $\Exp[|\tilde{v}_1^k-v_1^k|]=\Exp[|\tilde{q}_1^k-q_1^k|]<\frac{\phi_k}{2^{r_V}}$ from the quantizer design in LT-DMLE.} Accordingly, we rewrite \eqref{eq:thm_aw4_1} as
    \be
        \label{eq:thm_aw4_2}
        \sqrt{U_{t_{\cI}}} \Exp[|\tilde{x}_{t_{\cI}}-\hat{x}_{t_{\cI}}|] < O(1) \Big( \sum_{k=1}^K \frac{\Theta(\sqrt{U_{t_{\cI}}}/d_k)}{2^{r_V}} + \sum_{k=1}^K \frac{d_k}{\Theta(\sqrt{U_{t_{\cI}}})} \Big),
    \ee
    which concludes the proof since $U_{t_{\cI}}=\cI+O(1)$ as shown in the proof of Theorem \ref{thm:awgn1}.

\section*{Appendix E: Proof of Lemma \ref{lem:appA}}

    We start with the level-triggered sampling based estimators, and continue with the uniform sampling based ones.
    \textcolor{blue}{Since the quantized overshoot $\tilde{p}_n^k$ is between the smallest and largest quantization levels, the quantized incremental process $\tilde{u}_n^k$ lies in the interval $[e_k+\frac{\theta_k}{2^{r_U+1}},e_k+\theta_k\frac{2^{r_U+1}-1}{2^{r_U+1}}]$. Note that $\tilde{U}_{\tilde{\cT}}$ cannot exceed the target Fisher information, $\cI$, by more than $\sum_{k=1}^K (e_k+\theta_k\frac{2^{r_U+1}-1}{2^{r_U+1}})$, in which case all sensors transmit their largest possible messages at the stopping time. Hence, we write
    \be
        \label{eq:appA_1}
        \cI \leq \sum_{k=1}^K \sum_{n=1}^{M_{\tilde{\cT}}^k} \tilde{u}_n^k < \cI+\sum_{k=1}^K \left(e_k+\theta_k\frac{2^{r_U+1}-1}{2^{r_U+1}}\right),
    \ee
    followed by
    \begin{align}
    \begin{split}
        \label{eq:appA_2}
        \frac{\cI}{\tilde{\cT}}<\sum_{k=1}^K \left(e_k+\theta_k\frac{2^{r_U+1}-1}{2^{r_U+1}}\right) \frac{M_{\tilde{\cT}}^k}{\tilde{\cT}}, \\
        \text{and} \ \ \
        \frac{\cI+\sum_{k=1}^K \left(e_k+\theta_k\frac{2^{r_U+1}-1}{2^{r_U+1}}\right)}{\tilde{\cT}}>\sum_{k=1}^K e_k\frac{M_{\tilde{\cT}}^k}{\tilde{\cT}}.
        \end{split}
    \end{align}
Since from \eqref{eq:Ustop} $\tilde{\cT}\to\infty$ as $\cI\to\infty$, we have $\frac{M_{\tilde{\cT}}^k}{\tilde{\cT}}\to\frac{1}{\Exp[t_{1,U}^k]}$ by the the strong law of large numbers for renewal processes \cite[Proposition 3.3.1]{Ross96}. Using (A2) we can show that $\Exp[t_{1,U}^k]=\Theta(e_k)$ in the same way it was shown in the proof of Theorem \ref{thm:awgn4} that $\Exp[t_{1,V}^k]=\Theta(d_k)$. Hence, as $\cI\to\infty$ we rewrite \eqref{eq:appA_2} as
\begin{align}
    \begin{split}
        \label{eq:appA_3}
        \frac{\tilde{\cT}}{\cI}>\frac{1}{O(1)+\sum_{k=1}^K \frac{O(1)}{\Theta(e_k)}}, \\
        \text{and} \ \ \
        \frac{\tilde{\cT}}{\cI+\sum_{k=1}^K e_k+O(1)}<O(1),
        \end{split}
    \end{align}
        from which it is seen that $\tilde{\cT}=\Theta(\cI)$ if either $e_k=O(1)$ or $e_k\to\infty$ with $e_k=O(\cI)$, concluding the proof for LT-sDMLE and LT-dsDMLE.}

  \textcolor{blue}{  Note that, for U-sDMLE and U-dsDMLE, when the scheme stops at time $\tilde{\cT}$, the overshoot over $\cI$ is upper bounded by the sum of the largest quantization levels $\sum_{k=1}^K T_U\theta_k\frac{2^{r_U+1}-1}{2^{r_U+1}}$ of $\{\tilde{u}_{mT_U}^k\}_k$, corresponding to the worst case scenario in which $\tilde{U}_{(M_{\tilde{\cT}}-1)T_U}$ is just below $\cI$, and all sensors transmit the largest message possible at time $\tilde{\cT}$. Here, $M_{\tilde{\cT}} = \tilde{\cT}/T_U$ is the number of sampling times until $\tilde{\cT}$.
        Hence, similar to \eqref{eq:appA_1} we write
        \be
            \label{eq:appA_8}
            \cI \leq \sum_{k=1}^K \sum_{m=1}^{M_{\tilde{\cT}}} \tilde{u}_{mT_U}^k < \cI+\sum_{k=1}^K T_U\theta_k\frac{2^{r_U+1}-1}{2^{r_U+1}}.
        \ee
        Since $\frac{T_U\theta_k}{2^{r_U+1}}<\tilde{u}_{mT_U}^k<T_U\theta_k\frac{2^{r_U+1}-1}{2^{r_U+1}}$, we have
        \be
            \label{eq:appA_9}
            \frac{\cI}{T_U \sum_{k=1}^K\theta_k} \frac{2^{r_U+1}}{2^{r_U+1}-1} < M_{\tilde{\cT}} < \frac{\cI~2^{r_U+1}}{T_U \sum_{k=1}^K\theta_k} + 2^{r_U+1}-1,
        \ee
        and thus
        \be
            \label{eq:appA_10}
            \frac{\cI}{\sum_{k=1}^K\theta_k} \frac{2^{r_U+1}}{2^{r_U+1}-1} < \tilde{\cT} < \frac{\cI~2^{r_U+1}}{\sum_{k=1}^K\theta_k} + T_U(2^{r_U+1}-1).
        \ee
        Hence, $\tilde{\cT}=\Theta(\cI)$ if $T_U=O(\cI)$, concluding the proof.}

\section*{Appendix F: Proof of Lemma \ref{lem:appB}}

    We present the proofs first for LT-sDMLE and LT-dsDMLE, and then for U-sDMLE and U-dsDMLE.
    Note that we have $U_{\tilde{\cT}}^k = \sum_{n=1}^{M_{\tilde{\cT}}^k} u_n^k + \sum_{\tau=t_{M_{\tilde{\cT}}^k,U}^k+1}^{\tilde{\cT}} \frac{2|h^k_{\tau}|^2}{\sigma_k^2}$, and from \eqref{eq:URecov} that $\tilde{U}_{\tilde{\cT}}^k = \sum_{n=1}^{M_{\tilde{\cT}}^k} \tilde{u}_n^k$.
    Hence, we write
    \be
        \label{eq:appB_1}
        |U_{\tilde{\cT}}-\tilde{U}_{\tilde{\cT}}| \leq \sum_{k=1}^K \Bigg(\sum_{n=1}^{M_{\tilde{\cT}}^k}|\tilde{u}_n^k-u_n^k| + \sum_{\tau=t_{M_{\tilde{\cT}}^k,U}^k+1}^{\tilde{\cT}} \frac{2|h^k_{\tau}|^2}{\sigma_k^2}\Bigg),
    \ee
    where $\sum_{\tau=t_{M_{\tilde{\cT}}^k,U}^k+1}^{\tilde{\cT}} \frac{2|h^k_{\tau}|^2}{\sigma_k^2} < e_k$ since no sampling occurs between $t_{M_{\tilde{\cT}}^k,U}^k$ and $\tilde{\cT}$. \textcolor{blue}{Similar to \eqref{eq:ren_th} using \cite[Theorem 3.6.1]{Ross96} we can write $\frac{\sum_{n=1}^{M_{\tilde{\cT}}^k}|\tilde{u}_n^k-u_n^k|}{\tilde{\cT}}\to\frac{\Exp[|\tilde{u}_1^k-u_1^k|]}{\Exp[t_{1,U}^k]}$ since $\tilde{\cT}\to\infty$ as $\cI\to\infty$ from \eqref{eq:Ustop}. We have $\Exp[|\tilde{u}_1^k-u_1^k|]<\frac{\theta_k}{2^{r_U+1}}=O(1)$ due to the quantizer design, and $\Exp[t_{1,U}^k]=\Theta(e_k)$ from the proof of Lemma \ref{lem:appA}. Hence, as $\cI\to\infty$ \eqref{eq:appB_1} becomes
    \be
        \frac{|U_{\tilde{\cT}}-\tilde{U}_{\tilde{\cT}}|}{\cI} < \sum_{k=1}^K \left( \frac{\tilde{\cT}~O(1)}{\cI~\Theta(e_k)} + \frac{e_k}{\cI} \right),
    \ee
    where if $e_k\to\infty$ such that $e_k=o(\cI)$, $\tilde{\cT}=\Theta(\cI)$ as shown in Appendix E, and the right hand-side tends to zero.}

\textcolor{blue}{    For U-sDMLE and U-dsDMLE, similar to \eqref{eq:appB_1} we write
    \be
        \label{eq:appB_2}
        |\tilde{U}_{\tilde{\cT}}-U_{\tilde{\cT}}| \leq \sum_{k=1}^K \sum_{m=1}^{M_{\tilde{\cT}}}|\tilde{u}_{mT_U}^k-u_{mT_U}^k|,
    \ee
    where we lack the term representing the missing information at the FC between the last sampling time and the stopping time, e.g., the second term in \eqref{eq:appB_1}, since $\tilde{\cT}=M_{\tilde{\cT}}T_U$.
    It follows by the strong law of large numbers that $\frac{\sum_{m=1}^{M_{\tilde{\cT}}}|\tilde{u}_{mT_U}^k-u_{mT_U}^k|}{M_{\tilde{\cT}}}\to\Exp[|\tilde{u}_{T_U}^k-u_{T_U}^k|]$ as $\tilde{\cT}\to\infty$, i.e., $\cI\to\infty$. Note that $\Exp[|\tilde{u}_{T_U}^k-u_{T_U}^k|]<\frac{T_U\theta_k}{2^{r_U+1}}$ due to the quantizer design and $M_{\tilde{\cT}}=\frac{\tilde{\cT}}{T_U}$, hence as $\cI\to\infty$ using \eqref{eq:appB_2} we write
    \be
    \label{eq:appB_2n}
        \frac{|\tilde{U}_{\tilde{\cT}}-U_{\tilde{\cT}}|}{\cI} < \frac{\tilde{\cT}}{\cI}\sum_{k=1}^K \frac{\theta_k}{2^{r_U+1}},
    \ee
    where with $T_U=O(\cI)$ (e.g., constant $T_U$) $\cT=\Theta(\cI)$ from Lemma \ref{lem:appA}, and $|\tilde{U}_{\tilde{\cT}}-U_{\tilde{\cT}}|=o(\cI)$ if $r_U\to\infty$, concluding the proof.}
    \ignore{
    Next we rewrite the right hand-side of \eqref{eq:appB_2} as
    \begin{align}
    \label{eq:appB_3}
    	\sum_{k=1}^K \sum_{m=1}^{\infty} \bExp\Big[|\tilde{u}_{mT_U}^k-u_{mT_U}^k| \ind{M_{\tilde{\cT}}\geq m}\Big] =& \sum_{k=1}^K \sum_{m=1}^{\infty} \bExp\Big[|\tilde{u}_{mT_U}^k-u_{mT_U}^k| \Big] \bExp\big[\ind{M_{\tilde{\cT}}\geq m}\big] \\
	\label{eq:appB_4}
	<& \sum_{k=1}^K \frac{T_U\theta_k}{2^{r_U+1}} ~\bExp\big[M_{\tilde{\cT}}\big],
    \end{align}
    where to write \eqref{eq:appB_3} we used the fact that the event $M_{\tilde{\cT}}\geq m$ depends on the first $m-1$ messages from the sensors, i.e., $\{\tilde{u}_{nT_U}^k: n=1,\ldots,m-1; k=1,\ldots,K\}$, hence it is independent from the $m$-th message and the incremental process generated it. Note that \eqref{eq:appB_4} follows from $\bExp\Big[|\tilde{u}_{mT_U}^k-u_{mT_U}^k| \Big]<\frac{T_U\theta_k}{2^{r_U+1}}$, which is due to the quantizer design. Since $\{\sum_{k=1}^K \tilde{u}_{mT_U}^k\}_m$ are i.i.d., using Wald's identity we write
    \be
    \label{eq:appB_5}
    	\bExp\big[M_{\tilde{\cT}}\big]=\frac{\bExp[\tilde{U}_{\tilde{\cT}}]}{\bExp[\sum_{k=1}^K\tilde{u}_{T_U}^k]} < \frac{\cI+\sum_{k=1}^K T_U\theta_k \frac{2^{r_U+1}-1}{2^{r_U+1}}}{\sum_{k=1}^K c_k T_U\theta_k},
    \ee
    where we used the upper bound of $\tilde{U}_{\tilde{\cT}}$ (cf. Appendix E) and the average value of $\tilde{u}_{T_U}^k$ with $c_k\in(\frac{1}{2^{r_U+1}},\frac{2^{r_U+1}-1}{2^{r_U+1}})$ being some constant. Substituting \eqref{eq:appB_4} and \eqref{eq:appB_5} into \eqref{eq:appB_2} we can write
    \be
    \label{eq:appB_6}
    	\frac{\bExp\big[|\tilde{U}_{\tilde{\cT}}-U_{\tilde{\cT}}|\big]}{\cI} < \sum_{k=1}^K \frac{\cI ~T_U\theta_k+T_U\theta_k\sum_{k=1}^K T_U\theta_k \frac{2^{r_U+1}-1}{2^{r_U+1}}}{\cI ~2^{r_U+1} \sum_{k=1}^K c_k T_U\theta_k},
    \ee
    from which we conclude that $\bExp\big[|\tilde{U}_{\tilde{\cT}}-U_{\tilde{\cT}}|\big]=o(\cI)$ if $r_U\to\infty$ at any rate.
    }

\section*{Appendix G: Proof of Lemma \ref{lem:appC}}

    We again start with the level-triggered-sampling-based estimators, then continue with the uniform-sampling-based ones.
    Using the Cauchy-Schwarz inequality we write
    \begin{align}
        \frac{\bExp[|S_{\tilde{\cT}}|]}{\cI} &\leq \frac{\sqrt{\bExp[S_{\tilde{\cT}}^2]}}{\cI} = \frac{\sqrt{\bExp[U_{\tilde{\cT}}]}}{\cI} \nn\\
        \label{eq:appC_2}
        &< \sqrt{ \frac{ \tilde{U}_{\tilde{\cT}} + \sum_{k=1}^K \sum_{n=1}^{M_{\tilde{\cT}}^k}|\tilde{u}_n^k-u_n^k| + \sum_{k=1}^K e_k } {\cI^2} },
    \end{align}
    where we used \eqref{eq:appB_1} and the facts that $\bExp[U_{\tilde{\cT}}]=U_{\tilde{\cT}}$ [cf. \eqref{eq:Finf} and \eqref{eq:Ustop}] and $\sum_{\tau=t_{M_{\tilde{\cT}}^k,U}^k+1}^{\tilde{\cT}} \frac{2|h^k_{\tau}|^2}{\sigma_k^2} < e_k$ (cf. Appendix F) to write \eqref{eq:appC_2}.
    \textcolor{blue}{We showed in the proof of Lemma \ref{lem:appA} that $\tilde{U}_{\tilde{\cT}}$ is bounded by $\cI+\sum_{k=1}^K \left(e_k+\theta_k\frac{2^{r_U+1}-1}{2^{r_U+1}}\right)$. Since $\sum_{k=1}^K \sum_{n=1}^{M_{\tilde{\cT}}^k}|\tilde{u}_n^k-u_n^k|<\tilde{\cT} \sum_{k=1}^K \frac{O(1)}{\Theta(e_k)}$ (cf. Appendix F), as $\cI\to\infty$, \eqref{eq:appC_2} becomes
    \be
    \label{eq:appC_3}
    	\frac{\bExp[|S_{\tilde{\cT}}|]}{\cI} < \sqrt{ \frac{\sum_{k=1}^K 2e_k}{\cI^2} + \frac{\tilde{\cT}}{\cI^2} \sum_{k=1}^K \frac{O(1)}{\Theta(e_k)} }.
    \ee
    From \eqref{eq:appA_3}, $\tilde{\cT}<\Theta(\cI+\sum_{k=1}^K e_k)$. Using this upper bound for $\tilde{\cT}$ in \eqref{eq:appC_3} we see that the second term on the right hand-side tends to zero provided that $\frac{1}{\cI e_k}\to0,\forall k$, i.e., $\frac{1}{e_k}=o(\cI),\forall k$, and $\frac{\sum_{k=1}^K e_k}{\cI^2 e_k}\to0,\forall k$, i.e., $\frac{\sum_{k=1}^K e_k}{e_k}=o(\cI^2),\forall k$. Since $e_k=o(\cI^2),\forall k$ due to the first term on the right hand-side of \eqref{eq:appC_3}, these conditions are met if $e_k\not=o(1)$, concluding the first part of the proof.}
    \ignore{
    From Lemma \ref{lem:appA}, if $e_k=O(\cI)$ such that $e_k\not=o(1)$, then $\tilde{\cT}=\Theta(\cI)$ and the right hand-side of \eqref{eq:appC_3} tends to zero, which suffices to prove the first part of Lemma \ref{lem:appC}. Although not needed for analysis purposes, we can find some weaker conditions under which the first part of Lemma \ref{lem:appC} still holds.
    From \eqref{eq:appA_3} it is seen that $\tilde{\cT}=o(\cI\sum_{k=1}^K e_k)$ if $e_k=\omega(\cI)$ for some $k$. Using this new upper bound for $\tilde{\cT}$ in \eqref{eq:appC_3} we see that the second term on the right hand-side tends to zero provided that $\frac{\sum_{k=1}^K e_k}{e_k}=O(\cI),\forall k$, i.e., $\frac{e_j}{e_k}=O(\cI),\forall j,k$. It says that there should not be significant differences between the growth rates of the thresholds $\{e_k\}$. For instance, ensuring the same rate for all $e_k$, i.e., $e_k=\Theta(e),\forall k$, is the easiest way to satisfy that condition. Then, considering the constraint $e_k=o(\cI^2)$ due to the first term on the right hand-side of \eqref{eq:appC_3} we can relax the constraint in the first part of Lemma \ref{lem:appC} as $e_k=o(\cI^2)$ and $e_k\not=o(1)$ such that $\frac{e_j}{e_k}=O(\cI),\forall j,k$.

    For the second term on the numerator we write
    \begin{align}
    \label{eq:appC_n1}
    	\sum_{k=1}^K \bExp\Big[\sum_{n=1}^{M_{\tilde{\cT}}^k}|\tilde{u}_n^k-u_n^k|\Big] <& \sum_{k=1}^K \bExp\Big[\sum_{n=1}^{M_{\tilde{\cT}}^k+1}|\tilde{u}_n^k-u_n^k|\Big] \nn\\
	=& \sum_{k=1}^K (\bExp[M_{\tilde{\cT}}^k]+1) \bExp[|\tilde{u}_1^k-u_1^k|],
    \end{align}
    where we used Wald's identity for the i.i.d. process $\{|\tilde{u}_n^k-u_n^k|\}_n$ and the stopping time $M_{\tilde{\cT}}^k+1$. We should note here that $M_{\tilde{\cT}}^k$ is not a stopping time for $\{|\tilde{u}_n^k-u_n^k|\}_n$, that is why the above inequality was written. For a detailed discussion on this technical point we refer to \cite[Section IV-B]{Yilmaz13}. From the quantizer design $\bExp[|\tilde{u}_1^k-u_1^k|]<\frac{\theta_k}{2^{r_U+1}}$. Using again Wald's identity for the i.i.d. process $\{\tilde{u}_n^k\}_n$ and the stopping time $M_{\tilde{\cT}}^k+1$ we have $M_{\tilde{\cT}}^k+1=\frac{\bExp\Big[\sum_{n=1}^{M_{\tilde{\cT}}^k+1}\tilde{u}_n^k\Big]}{\bExp[\tilde{u}_1^k]}$, where $\bExp[\tilde{u}_1^k]>e_k+\frac{\theta_k}{2^{r_U+1}}$ by the quantizer design. Hence, \eqref{eq:appC_n1} can be rewritten as
    \be
    \label{eq:appC_n2}
    	\sum_{k=1}^K \bExp\Big[\sum_{n=1}^{M_{\tilde{\cT}}^k}|\tilde{u}_n^k-u_n^k|\Big]  < \frac{\max_k \frac{\theta_k}{2^{r_U+1}}}{\min_k \left(e_k+\frac{\theta_k}{2^{r_U+1}}\right)} \bExp\Big[\sum_{k=1}^K \sum_{n=1}^{M_{\tilde{\cT}}^k+1}\tilde{u}_n^k\Big],
    \ee
    where $\bExp\Big[\sum_{k=1}^K \sum_{n=1}^{M_{\tilde{\cT}}^k+1}\tilde{u}_n^k\Big]=\bExp\Big[\sum_{k=1}^K \sum_{n=1}^{M_{\tilde{\cT}}^k}\tilde{u}_n^k\Big]+\bExp\Big[\sum_{k=1}^K \tilde{u}_{M_{\tilde{\cT}}^k+1}^k\Big]$. Using the inequalities $\sum_{k=1}^K \sum_{n=1}^{M_{\tilde{\cT}}^k} \tilde{u}_n^k < \cI+\sum_{k=1}^K \left(e_k+\theta_k\frac{2^{r_U+1}-1}{2^{r_U+1}}\right)$ from \eqref{eq:appA_1} and $\tilde{u}_{M_{\tilde{\cT}}^k+1}^k<e_k+\frac{\theta_k}{2^{r_U+1}}$ by the quantizer design in \eqref{eq:appC_n2}, and then substituting \eqref{eq:appC_n2} in \eqref{eq:appC_2}, as $\cI\to\infty$ we write
    \be
    \label{eq:appC_n3}
    	\frac{\bExp[|S_{\tilde{\cT}}|]}{\cI} < \sqrt{\frac{\sum_{k=1}^K 2e_k }{\cI^2} + \frac{\max_k \frac{\theta_k}{2^{r_U+1}}}{\min_k \left(e_k+\frac{\theta_k}{2^{r_U+1}}\right)} \frac{\cI+\sum_{k=1}^K 2\left(e_k+\theta_k\frac{2^{r_U+1}-1}{2^{r_U+1}}\right)}{\cI^2}}.
    \ee
    With $e_k=o(\cI^2),\forall k$, as $\cI\to\infty$, the right hand-side of \eqref{eq:appC_n3} tends to zero, and thus $\bExp[|S_{\tilde{\cT}}|]=o(\cI)$.
    }

\textcolor{blue}{For U-sDMLE and U-dsDMLE, similar to \eqref{eq:appC_2} we write
\begin{align}
	\frac{\bExp[|S_{\tilde{\cT}}|]}{\cI} \leq& \sqrt{ \frac{ \tilde{U}_{\tilde{\cT}} + |\tilde{U}_{\tilde{\cT}}-U_{\tilde{\cT}}| } {\cI^2} } \nn\\
	\label{eq:appC_4}
	<& \sqrt{ \frac{T_U}{\cI^2}O(1) + \frac{\tilde{\cT}}{\cI^2}O(1) },
\end{align}
which follows from \eqref{eq:appA_8} and \eqref{eq:appB_2n}. If $T_U=o(\cI^2)$, we conclude that $\tilde{\cT}=o(\cI^2)$ from \eqref{eq:appA_10}, and accordingly the right hand-side of \eqref{eq:appC_4} tends to zero as $\cI\to\infty$.}

\section*{Appendix H: Proof of Theorem \ref{thm:fade1}}

    As in Theorem \ref{thm:awgn1} and Theorem \ref{thm:awgn3}, we will show convergence in mean, i.e, $\bExp[|\tilde{x}_{\tilde{\cT}}-x|]\to0$, to prove the theorem. Note that we can write $\tilde{x}_{\tilde{\cT}}-x$ as
    \be
        \label{eq:thm_fa1_1}
        \tilde{x}_{\tilde{\cT}}-x = \frac{U_{\tilde{\cT}}}{\tilde{U}_{\tilde{\cT}}} \left( \frac{\tilde{V}_{\tilde{\cT}}}{U_{\tilde{\cT}}}-\frac{\tilde{U}_{\tilde{\cT}}}{U_{\tilde{\cT}}} x \right).
    \ee
    Now, as we did before in Theorem \ref{thm:awgn1} and Theorem \ref{thm:awgn3}, we add and subtract $\hat{x}_{\tilde{\cT}}$ inside the parentheses, i.e., $\tilde{x}_{\tilde{\cT}}-x = \frac{U_{\tilde{\cT}}}{\tilde{U}_{\tilde{\cT}}} \left( \frac{\tilde{V}_{\tilde{\cT}}}{U_{\tilde{\cT}}}- \hat{x}_{\tilde{\cT}} + \hat{x}_{\tilde{\cT}} -\frac{\tilde{U}_{\tilde{\cT}}}{U_{\tilde{\cT}}} x \right)$. Replace the first $\hat{x}_{\tilde{\cT}}$ with $\frac{V_{\tilde{\cT}}}{U_{\tilde{\cT}}}$, and the second one with $x+\frac{S_{\tilde{\cT}}}{U_{\tilde{\cT}}}$. After distributing $\frac{U_{\tilde{\cT}}}{\tilde{U}_{\tilde{\cT}}}$ through the parentheses, and taking the absolute value and the expectation of both sides we get
    \be
        \label{eq:thm_fa1_2}
        \bExp\big[|\tilde{x}_{\tilde{\cT}}-x|\big] \leq \frac{\bExp\big[|\tilde{V}_{\tilde{\cT}}-V_{\tilde{\cT}}|\big]} {\cI} + \frac{|U_{\tilde{\cT}}-\tilde{U}_{\tilde{\cT}}|} {\cI} |x| + \frac{\bExp\big[|S_{\tilde{\cT}}|\big]}{\cI},
    \ee
    since $\tilde{U}_{\tilde{\cT}}\geq\cI$. If $e_k\to\infty$ such that $e_k=o(\cI),~\forall k$, the second term on the right hand-side of \eqref{eq:thm_fa1_2} tends to zero, following from Lemma \ref{lem:appB} and (A1), and similarly, the last term tends to zero, following from Lemma \ref{lem:appC}. \textcolor{blue}{For the first term we write $\frac{\bExp[|\tilde{V}_{\tilde{\cT}}-V_{\tilde{\cT}}|]}{\cI} \leq \sum_{k=1}^K \frac{\tilde{\cT}\phi_k}{\cI 2^{R_k}}$ due to the quantizer design}. Since $\tilde{\cT}=\Theta(\cI)$ from Lemma \ref{lem:appA}, the first term on the right hand-side of \eqref{eq:thm_fa1_2} tends to zero if $R_k\to\infty,\forall k$, at any rate as $\cI\to\infty$, concluding the proof.

\section*{Appendix I: Proof of Theorem \ref{thm:fade2}}

        Since $I_{\tilde{\cT}}^c=U_{\tilde{\cT}}$, from the proof of Theorem \ref{thm:fade1} we can write
    \be
        \label{eq:thm_fa2_1}
        \sqrt{U_{\tilde{\cT}}} (\tilde{x}_{\tilde{\cT}}-x) = \frac{U_{\tilde{\cT}}}{\tilde{U}_{\tilde{\cT}}} \Bigg( \frac{\tilde{V}_{\tilde{\cT}}-V_{\tilde{\cT}}} {\sqrt{U_{\tilde{\cT}}}} + \frac{U_{\tilde{\cT}}-\tilde{U}_{\tilde{\cT}}} {\sqrt{U_{\tilde{\cT}}}}~x + \frac{S_{\tilde{\cT}}}{\sqrt{U_{\tilde{\cT}}}} \Bigg).
    \ee
    Note from Section \ref{sec:optCen} that $\frac{S_{\tilde{\cT}}}{\sqrt{U_{\tilde{\cT}}}} \sim \cN(0,1)$. Hence, it is sufficient to show that $\frac{U_{\tilde{\cT}}}{\tilde{U}_{\tilde{\cT}}} \to 1$, $\frac{\tilde{V}_{\tilde{\cT}}-V_{\tilde{\cT}}} {\sqrt{U_{\tilde{\cT}}}} \to 0$, and $\frac{U_{\tilde{\cT}}-\tilde{U}_{\tilde{\cT}}} {\sqrt{U_{\tilde{\cT}}}} \to 0$ as $\cI\to\infty$. \textcolor{blue}{It is shown in \eqref{eq:appA_1} that $\cI \leq \tilde{U}_{\tilde{\cT}} < \cI+\sum_{k=1}^K \left(e_k+\theta_k\frac{2^{r_U+1}-1}{2^{r_U+1}}\right)$. For $U_{\tilde{\cT}}$, from \eqref{eq:appB_1} and the discussion after it we can write
    \be
        \label{eq:thm_fa2_2}
        \tilde{U}_{\tilde{\cT}} - \sum_{k=1}^K \Bigg(\sum_{n=1}^{M_{\tilde{\cT}}^k}|\tilde{u}_n^k-u_n^k| + e_k\Bigg) < U_{\tilde{\cT}} < \tilde{U}_{\tilde{\cT}} + \sum_{k=1}^K \Bigg(\sum_{n=1}^{M_{\tilde{\cT}}^k}|\tilde{u}_n^k-u_n^k| + e_k\Bigg),
    \ee
    where, by \cite[Theorem 3.6.1]{Ross96}, $\frac{\sum_{n=1}^{M_{\tilde{\cT}}^k}|\tilde{u}_n^k-u_n^k|}{\tilde{\cT}}\to\frac{\Exp[|\tilde{u}_1^k-u_1^k|]}{\Exp[t_{1,U}^k]}<\frac{\theta_k}{\Theta(e_k)2^{r_U+1}}$ as $\cI\to\infty$.
    Using the lower and upper bounds for $\tilde{U}_{\tilde{\cT}}$ and the fact that $\tilde{\cT}=\Theta(\cI)$ from Appendix E, as $\cI\to\infty$, we can write
    \be
    \label{eq:thm_fa2_2n}
        \underbrace{ \frac{\cI-\sum_{k=1}^K\left[e_k+\Theta\left(\frac{\cI}{e_k}\right)\right]}{\cI+\sum_{k=1}^K e_k+O(1)} }_{1-\frac{\sum_{k=1}^K\left[\Theta(e_k)+\Theta\left(\frac{\cI}{e_k}\right)\right]+O(1)}{\cI+\sum_{k=1}^K e_k+O(1)}} < \frac{U_{\tilde{\cT}}}{\tilde{U}_{\tilde{\cT}}} < \underbrace{ \frac{\cI+\sum_{k=1}^K\left[\Theta(e_k)+\Theta\left(\frac{\cI}{e_k}\right)\right]+O(1)}{\cI} }_{1+\frac{\sum_{k=1}^K\left[\Theta(e_k)+\Theta\left(\frac{\cI}{e_k}\right)\right]+O(1)}{\cI}},
    \ee
    hence $\frac{U_{\tilde{\cT}}}{\tilde{U}_{\tilde{\cT}}} \to 1$ if $e_k\to\infty$ such that $e_k=o(\cI)$.
    From the quantizer design we have $\bExp[|\tilde{V}_{\tilde{\cT}}-V_{\tilde{\cT}}|]<\sum_{k=1}^K\frac{\tilde{\cT}\phi_k}{2^{R_k}}$, hence by \eqref{eq:thm_fa2_2}
    \be
    \label{eq:thm_fa2_3}
        \frac{\bExp[|\tilde{V}_{\tilde{\cT}}-V_{\tilde{\cT}}|]}{\sqrt{U_{\tilde{\cT}}}}<\sum_{k=1}^K\frac{\tilde{\cT}\phi_k}{2^{R_k}\sqrt{\cI-\sum_{k=1}^K \Big(\sum_{n=1}^{M_{\tilde{\cT}}^k}|\tilde{u}_n^k-u_n^k| + e_k\Big)}}.
    \ee
    With $e_k\to\infty$ such that $e_k=o(\cI)$, as $\cI\to\infty$, \eqref{eq:thm_fa2_3} becomes $\frac{\bExp[|\tilde{V}_{\tilde{\cT}}-V_{\tilde{\cT}}|]}{\sqrt{U_{\tilde{\cT}}}}<\sum_{k=1}^K\frac{\Theta(\sqrt{\cI})}{2^{R_k}}$ since $\sum_{n=1}^{M_{\tilde{\cT}}^k}|\tilde{u}_n^k-u_n^k|<\Theta\big(\frac{\tilde{\cT}}{e_k}\big)$ and $\tilde{\cT}=\Theta(\cI)$. Thus, $\frac{\bExp[|\tilde{V}_{\tilde{\cT}}-V_{\tilde{\cT}}|]}{\sqrt{U_{\tilde{\cT}}}}\to0$ if $R_k=\omega(\log \cI)$, implying, by Markov's inequality, $\frac{|\tilde{V}_{\tilde{\cT}}-V_{\tilde{\cT}}|}{\sqrt{U_{\tilde{\cT}}}} \to 0$, which in turn implies $\frac{\tilde{V}_{\tilde{\cT}}-V_{\tilde{\cT}}}{\sqrt{U_{\tilde{\cT}}}} \to 0$.
    Similarly, from \eqref{eq:appB_1} and the discussion after it we see that $\frac{|U_{\tilde{\cT}}-\tilde{U}_{\tilde{\cT}}|}{\sqrt{U_{\tilde{\cT}}}}<\frac{\Theta(\sqrt{\cI}/e_k)}{2^{r_U+1}}+\sum_{k=1}^K \frac{e_k}{\sqrt{\cI}}$, hence it tends to zero if $e_k=o(\sqrt{\cI})$ and $r_U=\omega(\log(\sqrt{\cI}/e_k)),~\forall k$, concluding the proof.}

\section*{Appendix J: Proof of Theorem \ref{thm:fade3}}

    LT-dsDMLE differs from LT-sDMLE only in the transmission of $\tilde{V}_{\tilde{\cT}}$, hence the proof of Theorem \ref{thm:fade1} up to and including \eqref{eq:thm_fa1_2} applies here. Moreover, as in Theorem \ref{thm:fade1}, if $e_k\to\infty$ such that $e_k=o(\cI),~\forall k$, we have $\frac{|U_{\tilde{\cT}}-\tilde{U}_{\tilde{\cT}}|}{\cI}\to0$, and $\frac{\bExp[|S_{\tilde{\cT}}|]}{\cI}\to0$ from Lemma \ref{lem:appB} and Lemma \ref{lem:appC}, respectively. For $\frac{\bExp[|\tilde{V}_{\tilde{\cT}}-V_{\tilde{\cT}}|]}{\cI}$, following \eqref{eq:thm_aw3_2}, as $\cI\to\infty$, \textcolor{blue}{we write
    \be
        \label{eq:thm_fa3_1}
        \frac{\bExp\big[|\tilde{V}_{\tilde{\cT}}-V_{\tilde{\cT}}|\big]}{\cI} < \sum_{k=1}^K \bExp\left[ \frac{\sum_{n=1}^{N_{\tilde{\cT}}^k} |\tilde{v}_n^k-v_n^k|}{N_{\tilde{\cT}}^k} \frac{N_{\tilde{\cT}}^k}{\cI} \right] + \sum_{k=1}^K \frac{d_k}{\cI}.
    \ee
    Since $\{|\tilde{v}_n^k-v_n^k|\}$ are independent and non-negative random variables, from \cite[Lemma 2]{Smith64} we write $\frac{\sum_{n=1}^{N_{\tilde{\cT}}^k} |\tilde{v}_n^k-v_n^k|}{N_{\tilde{\cT}}^k} \to \frac{\sum_{n=1}^{N_{\tilde{\cT}}^k} \bExp[|\tilde{v}_n^k-v_n^k|]}{N_{\tilde{\cT}}^k}<\frac{\phi_k}{2^{r_V}}$ as $N_{\tilde{\cT}}^k\to\infty$, i.e., as $\cI\to\infty$, the inequality being true due to the quantizer design. Using the elementary renewal theorem for non-identically distributed variables \cite[Theorem 2]{Smith64} we can write $\frac{\bExp[N_{\tilde{\cT}}^k]}{\tilde{\cT}} \to \frac{N_{\tilde{\cT}}^k}{\sum_{n=1}^{N_{\tilde{\cT}}^k} \bExp[s_n^k]}$, where $s_n^k \triangleq t_{n,V}^k-t_{n-1,V}^k$ is the $n$-th inter-sampling interval. Now consider a new sampling process $\{\bar{s}_n^k\}$ with the observation $\left| \frac{2\Re((y^k_t)^*h^k_t)}{\sigma_k^2} \right|$ at time $t$ instead of $\frac{2\Re((y^k_t)^*h^k_t)}{\sigma_k^2}$ [cf. \eqref{eq:samp}], and the threshold $d_k$. Note that $s_n^k \geq \bar{s}_n^k,\forall n,k$. By again \cite[Theorem 2]{Smith64} we have $\frac{\bExp[\bar{s}_n^k]}{d_k} \to \frac{\bar{s}_n^k}{\sum_{\tau=1}^{\bar{s}_n^k} \frac{2}{\sigma_k^2}\bExp\left[|\Re((y^k_t)^*h^k_t)|\right]}$ as $d_k\to\infty$, where $\bExp\left[|\Re((y^k_t)^*h^k_t)|\right]=O(1)$ by \eqref{eq:bdd_obs}. Hence, $\bExp[\bar{s}_n^k]=\Theta(d_k)$ and $\frac{\bExp[N_{\tilde{\cT}}^k]}{\tilde{\cT}}<\Theta\left(\frac{1}{d_k}\right)$.
    With $e_k\to\infty$ such that $e_k=o(\cI),\forall k$, from Lemma \ref{lem:appA}, $\tilde{\cT}=\Theta(\cI)$, thus the right hand-side of \eqref{eq:thm_fa3_1} tends to zero if $d_k\to\infty$ such that $d_k=o(\cI),\forall k$, concluding the proof.}
    \ignore{
    \begin{align}
        \label{eq:thm_fa3_1}
        \frac{\bExp\big[|\tilde{V}_{\tilde{\cT}}-V_{\tilde{\cT}}|\big]}{\cI} &\leq \frac{\sum_{k=1}^K \bExp\Big[\sum_{n=1}^{N_{\tilde{\cT}}^k}|\tilde{v}_n^k-v_n^k|\Big]}{\cI} + \frac{\sum_{k=1}^K \bExp\Big[\big|\sum_{\tau=t_{N_{\tilde{\cT}}^k,V}^k+1}^{\tilde{\cT}} \frac{2\Re((y^k_{\tau})^*h^k_{\tau})}{\sigma_k^2}\big|\Big]}{\cI} \nn\\
        &< \frac{\phi}{2^{r_V-1}} \sum_{k=1}^K \frac{\bExp\big[N_{\tilde{\cT}}^k\big]}{\cI} + \sum_{k=1}^K \frac{d_k}{\cI},
    \end{align}
    where we used $|\tilde{v}_n^k-v_n^k| < \phi/2^{r_V-1}$ (cf. proof of Theorem \ref{thm:awgn4}), and $\Big|\sum_{\tau=t_{N_{\tilde{\cT}}^k,V}^k+1}^{\tilde{\cT}} \frac{2\Re((y^k_{\tau})^*h^k_{\tau})}{\sigma_k^2}\Big| < d_k$ (cf. proof of Theorem \ref{thm:awgn3}). The second term on the right hand-side of \eqref{eq:thm_fa3_1} tends to zero if $d_k=o(\cI),~\forall k$.

    Note that the shortest intersampling interval, $\tau_{n,V}^k \triangleq t_{n,V}^k-t_{n-1,V}^k$, happens when $\left\{\frac{2\Re((y^k_{\tau})^*h^k_{\tau})}{\sigma_k^2}\right\},~\tau \in (t_{n-1,V}^k,t_{n,V}^k]$, have the same sign and the largest possible magnitude, which is less than $\phi$, i.e., $\tau_{n,V}^k > \frac{d_k}{\phi},~\forall n,k$. Assume that $\tau_{n,V}^k < \infty,~\forall n,k$, and there exists a constant $C$ such that $\tau_{n,V}^k < C \frac{d_k}{\phi},~\forall n,k$, where $1<C<\infty$. Using these bounds on $\tau_{n,V}^k$ we can write
    \begin{align}
        \label{eq:thm_fa3_2}
        \sum_{n=1}^{N_{\tilde{\cT}}^k} \tau_{n,V}^k &\leq \tilde{\cT} < \sum_{n=1}^{N_{\tilde{\cT}}^k+1} \tau_{n,V}^k, \\
        \frac{d_k}{\phi} N_{\tilde{\cT}}^k &< \tilde{\cT} < C \frac{d_k}{\phi} (N_{\tilde{\cT}}^k+1), \nn\\
        \frac{\phi}{C d_k} \tilde{\cT} - 1 &< N_{\tilde{\cT}}^k < \frac{\phi}{d_k} \tilde{\cT}, \nn\\
        \label{eq:thm_fa3_3}
        \frac{\phi}{C d_k} \frac{\bExp\big[\tilde{\cT}\big]}{\cI} - \frac{1}{\cI} &< \frac{\bExp\big[N_{\tilde{\cT}}^k\big]}{\cI} < \frac{\phi}{d_k} \frac{\bExp\big[\tilde{\cT}\big]}{\cI}.
    \end{align}
    Since $\bExp\big[\tilde{\cT}\big]=\Theta(\cI)$ from Lemma \ref{lem:appA}, and $\frac{1}{\cI}=o\left(\frac{1}{d_k}\right)$ for $d_k=o(\cI)$, we have $\frac{\bExp\big[N_{\tilde{\cT}}^k\big]}{\cI}=\Theta\left(\frac{1}{d_k}\right)$, substituting which in \eqref{eq:thm_fa3_1} we conclude the proof.
    }

\section*{Appendix K: Proof of Theorem \ref{thm:fade4}}

    The proof follows that of Theorem \ref{thm:fade2}, except for the part showing $\frac{\tilde{V}_{\tilde{\cT}}-V_{\tilde{\cT}}}{\sqrt{U_{\tilde{\cT}}}} \to 0$ since LT-dsDMLE differs from LT-sDMLE only in the transmission of $\tilde{V}_{\tilde{\cT}}$. From Appendix J, we have $\frac{\bExp[|\tilde{V}_{\tilde{\cT}}-V_{\tilde{\cT}}|]}{\sqrt{\cI}} \to 0$ if $d_k=o(\sqrt{\cI})$ and $r_V=\omega(\log(\sqrt{\cI}/d_k))$, implying, by Markov's inequality, $\frac{|\tilde{V}_{\tilde{\cT}}-V_{\tilde{\cT}}|}{\sqrt{\cI}}\to0$, and thus $\frac{\tilde{V}_{\tilde{\cT}}-V_{\tilde{\cT}}}{\sqrt{\cI}}\to0$. We conclude the proof noting from \eqref{eq:thm_fa2_2n} that $U_{\tilde{\cT}}=\Theta(\cI)$ when $e_k\to\infty$ such that $e_k=o(\cI)$.

\section*{Appendix L: Proof of Theorem \ref{thm:uni1}}

    The first part of the proof follows from the proof of Theorem \ref{thm:fade1}, which uses Lemma \ref{lem:appA}, Lemma \ref{lem:appB} and Lemma \ref{lem:appC}.
    To show asymptotic optimality, we start with \eqref{eq:thm_fa2_1}. Similar to Theorem \ref{thm:fade2}, we will first show that $\frac{U_{\tilde{\cT}}}{\tilde{U}_{\tilde{\cT}}}\to1$. From \eqref{eq:appA_8}, \eqref{eq:appB_2} and \eqref{eq:appB_2n}, as $\cI\to\infty$,\textcolor{blue}{ we can write
    \be
    \label{eq:appUni1}
        1-\frac{T_U~O(1)+\tilde{\cT}~\frac{O(1)}{2^{r_U+1}}}{\cI+T_U~O(1)}<\frac{U_{\tilde{\cT}}}{\tilde{U}_{\tilde{\cT}}}<1+\frac{T_U~O(1)+\tilde{\cT}~\frac{O(1)}{2^{r_U+1}}}{\cI},
    \ee
    where with $T_U=o(\cI)$, from Lemma \ref{lem:appA}, $\tilde{\cT}=\Theta(\cI)$ and $\frac{U_{\tilde{\cT}}}{\tilde{U}_{\tilde{\cT}}}\to1$ if $r_U\to\infty$ at any rate.
    Now, we need to show that $\frac{\tilde{V}_{\tilde{\cT}}-V_{\tilde{\cT}}}{\sqrt{U_{\tilde{\cT}}}} \to 0$, and $\frac{U_{\tilde{\cT}}-\tilde{U}_{\tilde{\cT}}}{\sqrt{U_{\tilde{\cT}}}} \to 0$. Similar to \eqref{eq:thm_fa2_3} we can show that $\frac{\bExp[|\tilde{V}_{\tilde{\cT}}-V_{\tilde{\cT}}|]}{\sqrt{U_{\tilde{\cT}}}}<\sum_{k=1}^K\frac{\Theta(\sqrt{\cI})}{2^{R_k}}$, }hence $\frac{\bExp[|\tilde{V}_{\tilde{\cT}}-V_{\tilde{\cT}}|]}{\sqrt{U_{\tilde{\cT}}}} \to 0$ if $R_k=\omega(\log \cI)$, which implies $\frac{\tilde{V}_{\tilde{\cT}}-V_{\tilde{\cT}}}{\sqrt{U_{\tilde{\cT}}}} \to 0$ by Markov's inequality.
    Finally, using \eqref{eq:appB_2n} we show that $\frac{|U_{\tilde{\cT}}-\tilde{U}_{\tilde{\cT}}|}{\sqrt{U_{\tilde{\cT}}}} \to 0$, and thus $\frac{U_{\tilde{\cT}}-\tilde{U}_{\tilde{\cT}}}{\sqrt{U_{\tilde{\cT}}}} \to 0$ if $r_U=\omega(\log \cI)$.

\section*{Appendix M: Proof of Theorem \ref{thm:uni2}}

    Since U-dsDMLE differs from U-sDMLE in only the transmission of $V_{\tilde{\cT}}^k$, the proof follows that of Theorem \ref{thm:uni1} except for the part showing $\frac{\bExp\big[|\tilde{V}_{\tilde{\cT}}-V_{\tilde{\cT}}|\big]}{\cI}\to0$ and $\frac{\tilde{V}_{\tilde{\cT}}-V_{\tilde{\cT}}}{\sqrt{U_{\tilde{\cT}}}}\to0$. Similar to \eqref{eq:thm_fa3_1}, we write
    \be
        \label{eq:uni2_V}
        \frac{\bExp\big[|\tilde{V}_{\tilde{\cT}}-V_{\tilde{\cT}}|\big]}{\cI} \leq \frac{\sum_{k=1}^K \bExp\Big[\sum_{m=1}^{N_{\tilde{\cT}}}|\tilde{v}_{mT_V}^k-v_{mT_V}^k|\Big]}{\cI} + \frac{\sum_{k=1}^K \bExp\Big[\sum_{\tau=N_{\tilde{\cT}}T_V+1}^{\tilde{\cT}} \big|\frac{2\Re((y^k_{\tau})^*h^k_{\tau})}{\sigma_k^2}\big|\Big]}{\cI},
    \ee
    \textcolor{blue}{where $\bExp\Big[\sum_{\tau=N_{\tilde{\cT}}T_V+1}^{\tilde{\cT}} \big|\frac{2\Re((y^k_{\tau})^*h^k_{\tau})}{\sigma_k^2}\big|\Big]<T_V \bExp\Big[\big|\frac{2\Re((y^k_{\tau})^*h^k_{\tau})}{\sigma_k^2}\big|\Big]=\Theta(T_V)$ since $\tilde{\cT}-N_{\tilde{\cT}}T_V<T_V$, and $\bExp\big[|\Re((y^k_{\tau})^*h^k_{\tau})|\big]=O(1)$  by \eqref{eq:bdd_obs}. From the law of large numbers for i.n.i.d. and non-negative random variables \cite[Lemma 2]{Smith64} we write $\frac{\sum_{m=1}^{N_{\tilde{\cT}}}|\tilde{v}_{mT_V}^k-v_{mT_V}^k|}{N_{\tilde{\cT}}}\to\frac{\sum_{n=1}^{N_{\tilde{\cT}}} \bExp[|\tilde{v}_{T_V}^k-v_{T_V}^k|]}{N_{\tilde{\cT}}}<\frac{T_V\phi_k}{2^{r_V}}$ due to the quantizer design, as $\tilde{\cT}\to\infty$, i.e., $\cI\to\infty$. Note that $\frac{\tilde{\cT}}{T_V}\geq N_{\tilde{\cT}}$, thus we have $\bExp\Big[\sum_{m=1}^{N_{\tilde{\cT}}}|\tilde{v}_{mT_V}^k-v_{mT_V}^k|\Big]<\tilde{\cT}\frac{\phi_k}{2^{r_V}}$ as $\cI\to\infty$. Then, as $\cI\to\infty$, \eqref{eq:uni2_V} becomes
    \be
    	\frac{\bExp\big[|\tilde{V}_{\tilde{\cT}}-V_{\tilde{\cT}}|\big]}{\cI} < \frac{\tilde{\cT}}{\cI} \frac{\sum_{k=1}^K\phi_k}{2^{r_V}} + \frac{\Theta(T_V)}{\cI},
    \ee
where with $T_U=O(\cI)$, from Lemma \ref{lem:appA}, $\tilde{\cT}=\Theta(\cI)$, and thus the right hand-side tends to zero if $T_V=o(\cI)$ and $r_V\to\infty$ at any rate}. Similarly, $\frac{\bExp\big[|\tilde{V}_{\tilde{\cT}}-V_{\tilde{\cT}}|\big]}{\sqrt{\cI}}\to0$ if $r_V=\omega(\log \cI)$ and $T_V=o(\sqrt{\cI})$. We conclude the proof noting from \eqref{eq:appUni1} that $U_{\tilde{\cT}}=\Theta(\cI)$ and $\frac{\bExp\big[|\tilde{V}_{\tilde{\cT}}-V_{\tilde{\cT}}|\big]}{\sqrt{\cI}}\to0$ implies $\frac{\tilde{V}_{\tilde{\cT}}-V_{\tilde{\cT}}}{\sqrt{\cI}}\to0$ by Markov's inequality.

\section*{Acknowledgement}

The authors would like to thank Prof. George V. Moustakides for his valuable feedback.


\end{document}